\documentclass{article}

\usepackage{PRIMEarxiv}
\usepackage[utf8]{inputenc} 
\usepackage{amsmath,amsfonts,amssymb}
\usepackage{mathtools}
\usepackage{booktabs}
\usepackage{siunitx}
\usepackage{cite}
\usepackage{lettrine}
\usepackage[hidelinks]{hyperref}
\usepackage{comment}
\usepackage[caption=false,font=footnotesize]{subfig}
\usepackage{float}
\newfloat{algorithm}{t}{lop}
\usepackage{array}
\usepackage{booktabs}
\usepackage{url}
\usepackage{xcolor}
\usepackage{graphicx}
\usepackage{multirow}
\usepackage{makecell}
\usepackage{caption}
\usepackage[euler]{textgreek}
\usepackage{relsize}
\usepackage{amssymb}
\usepackage{bm}
\usepackage{dblfloatfix}
\usepackage[subnum]{cases}
\usepackage{algorithm}
\usepackage{algpseudocode}

\usepackage{physics}

\renewcommand{\mathbf}{\boldsymbol}

\usepackage{float}
\usepackage{capt-of}
\usepackage{tabularx}
\newcount\Comments  
\usepackage{color}
\definecolor{darkgreen}{rgb}{0,0.5,0}
\definecolor{purple}{rgb}{1,0,1}
\definecolor{amber}{rgb}{1,0.49,0}
\definecolor{uglycolor}{rgb}{0.5,0.5,0}
\newcommand{\kibitz}[2]{\ifnum\Comments=0\textcolor{#1}{#2}\fi}

\makeatletter
\newcommand*{\rom}[1]{\expandafter\@slowromancap\romannumeral #1@}
\makeatother

\title{Conceptualizing and Modeling Communication-Based Cyberattacks on Automated Vehicles
\thanks{\textit{\underline{Citation}}: 
\textbf{Li, Liu and Yang. Conceptualizing and Modeling Communication-Based Cyberattacks on Automated Vehicles.}} 
}

\author{
  Tianyi Li \\
  Department of Civil Engineering \\
  Saint Louis University \\
  \texttt{tianyili.ai@gmail.com} \\
  \AND
  Tianyu Liu \\
  School of Artificial Intelligence\\
  Hebei University of Technology  
  \AND
  Yicheng Yang \\
  School of Artificial Intelligence\\
  Hebei University of Technology  \\}

\begin{document}
\maketitle

\begin{abstract}
Adaptive Cruise Control (ACC) is rapidly proliferating across electric vehicles (EVs) and internal combustion engine (ICE) vehicles, enhancing traffic flow while simultaneously expanding the attack surface for communication-based cyberattacks. Because the two powertrains translate control inputs into motion differently, their cyber-resilience remains unquantified. 
Therefore, we formalize six novel message-level attack vectors and implement them in a ring-road simulation that systematically varies the ACC market penetration rates (MPRs) and the spatial pattern of compromised vehicles.  
A three-tier risk taxonomy converts disturbance metrics into actionable defense priorities for practitioners. Across all simulation scenarios, EV platoons exhibit lower velocity standard deviation, reduced spacing oscillations, and faster post-attack recovery compared to ICE counterparts, revealing an inherent stability advantage. 
These findings clarify how controller-to-powertrain coupling influences vulnerability and offer quantitative guidance for the detection and mitigation of attacks in mixed automated traffic.

\end{abstract}


\section{Introduction}\label{sec introduction}

\lettrine[lines=2]{A}{dvancements}  in vehicle automation have been widely recognized for their potential to significantly improve transportation systems. Automated Vehicles (AVs) can enhance traffic flow \cite{learn2017freeway}, facilitate effective coordination \cite{sun2022energy}, and contribute to greater energy efficiency \cite{ghanavati2017analysis}.
Moreover, AVs demonstrate the potential to significantly enhance roadway safety by reducing human error \cite{sun2021survey} and to contribute to environmental sustainability through optimized driving behaviors and reduced emissions \cite{lapardhaja2023unlocking}.

However, the increasing integration of adaptive cruise control (ACC)-equipped vehicles expands the potential vulnerability exposure, thereby heightening the risk of cyber threats. Malicious attacks targeting AVs can compromise operational safety, potentially resulting in traffic collisions, property damage, and threats to human life \cite{durlik2024cybersecurity,WANG2022100078}.
The inherent connectivity of AVs significantly increases their exposure to cyber threats. Given their reliance on inter-vehicle coordination and environmental sensing, AVs are especially susceptible to advanced attacks targeting automation functions.

Structurally, AVs consist of three primary subsystems: the driving system, the automotive control system, and the communication system \cite{li2018influence}, each of which presents distinct cybersecurity vulnerabilities.
Attacks on the driving system differ significantly from those on traditional vehicles without ACC systems \cite{kim2021cybersecurity}. 
Modern driving systems increasingly depend on a range of components, including the Global Positioning System (GPS), mobile applications, and sensors. And these sensors are composed of cameras, LiDAR \cite{el2020cybersecurity}, radar, and magnetic encoders \cite{kandefer2024asurvey}, all of which are used to receive and interpret external signals, enabling accurate perception and interaction with the surrounding environment. 
Sensors' constant interaction with dynamic environments renders them particularly susceptible to cyberattacks. For instance, GPS spoofing attacks poses a substantial threat to AVs due to their strong reliance on GPS signals for accurate localization and navigation \cite{chowdhury2020attacks}. The deliberate distortion or falsification of GPS data may induce erroneous position estimation, leading to incorrect decision-making and increasing the risk of traffic incidents.
%
In addition, jamming attacks specifically target the LiDAR sensor by emitting interfering signals, which obstruct the sensor’s ability to receive environmental information timely, thereby compromising the vehicle's situational awareness and operational safety \cite{el2020cybersecurity}.


Cyberattacks targeting the automotive control system often focus on critical components including the electronic control unit, inertial measurement unit (IMU) \cite{kim2021cybersecurity}, in-vehicle networks, automotive keys, and ACC system \cite{guembe2022emerging}. The aforementioned attacks may lead to leakage of personal data, degrade system performance, and impair inter-vehicle coordination. Notably, attacks on car keys can enable unauthorized vehicle ignition without the physical key \cite{cai2019roadways}; Distributed denial of service attacks targeting network interfaces can introduce significant delays in command transmission, disrupting timely vehicle operation \cite{chowdhury2020attacks}.

Cyberattacks targeting the communication system are often more frequent and pose greater risks due to their role as the essential conduit for real-time data exchange and inter-vehicle coordination \cite{durlik2024cybersecurity}.
The communication system facilitates Vehicle-to-Vehicle (V2V) \cite{kim2021cybersecurity}, Vehicle-to-Infrastructure (V2I) \cite{parkinson2017cyber}, and Vehicle-to-Pedestrian (V2P) interactions \cite{kumar2021black}. However, reliance on wireless communication inherently introduces security vulnerabilities that can be exploited by malicious actors \cite{lin2025fedav}.

%

%
Cyberattacks on V2V communication can disrupt the exchange of critical real-time information and coordination among connected vehicles. To name a few, spam attacks can consume the system bandwidth and degrade communication efficiency \cite{albouq2017lightweight};
Denial-of-Service (DoS) attacks can delay communication between vehicles, thereby affecting their normal driving behavior \cite{wang2023optimal};
Packet Dropping Attack (PDA) selectively discards data during message transmission \cite{edemacu2014packet}. Specifically, a malicious node masquerades as a router and drops incoming packets \cite{hassan2020intelligent}, disrupting data flow \cite{malik2022efficient} and compromising network security \cite{dhanaraj2021black}. 
Cyberattacks on V2I communication target the data transfer between vehicles and roadside units, enabling attackers to transmit false information \cite{pham2021survey}.
For instance, man-in-the-middle attacks intercept and alter communications between vehicles and external entities, compromising message integrity and potentially leading to unsafe or erroneous vehicle behavior \cite{wang2020modeling}.
Sybil attacks compromise infrastructure by transmitting fabricated identity data, misleading traffic control systems, and disrupting signal coordination, thereby degrading overall transportation efficiency \cite{pham2021survey}.
Replay attacks occur when an adversary retransmits recorded traffic signal information. It may cause vehicles to misinterpret current traffic conditions, leading to incorrect decision-making and compromised roadway safety \cite{kim2021cybersecurity}.

Cyberattacks on V2P communication target the transmission of critical safety information between vehicles and pedestrians, increasing the risk of accidents and compromising pedestrian safety \cite{sewalkar2019vehicle}.
For instance, false data injection attacks can disrupt communication systems by introducing erroneous information, preventing accurate reception of data from pedestrians \cite{wang2024cyberattacks}. Additionally, Sybil and DoS attacks can disrupt and delay information exchange between vehicles and pedestrians, thereby impairing V2P communication.
Therefore, cyberattacks on the communication system of AVs can burden network resources and may even render functional services unavailable \cite{li2021comprehensive,aslan2023comprehensive}.

With the rapid advancement of network technologies, AVs are increasingly capable of interacting with their surrounding environments. While traditional cyberattacks have been extensively studied, they are gradually becoming outdated as numerous variants emerge from these conventional attack types. Many of these evolving threats remain underexplored despite exhibiting greater intensity and disruptive potential, posing significant risks to the integrity and reliability of AV systems. Furthermore, the impact of cyberattacks varies considerably across different AV platforms.
To address these challenges, this study proposes six types of communication-based cyberattacks targeting AVs equipped with ACC, with particular emphasis on their effects on communication systems. We specifically investigate the differential responses between electric vehicles (EVs) and internal combustion engine (ICE) vehicles equipped with ACC when subjected to these attack scenarios. Through a comprehensive analysis of these cyberattacks, we aim to deepen understanding of their underlying mechanisms and consequences, thereby contributing to enhanced driving system safety and reliability.
The primary contributions and innovations of this study are presented as follows:

\begin{itemize} 
\item This study proposes and mathematically formalizes six novel communication-based cyberattacks targeting ACC-equipped vehicles. The analysis divides the driving process into three sequential phases: pre-attack, during-attack, and post-attack. By examining potential attack scenarios from the adversary’s perspective, this work provides critical insights into the underlying attack mechanisms and associated system vulnerabilities.

\item 
We successfully implement the proposed cyberattacks in a ring-road simulation that systematically varies the ACC market penetration rates (MPRs) and the spatial configurations of compromised vehicles, including non-adjacent and adjacent attacked patterns.

\item We conduct a comparative analysis to systematically assess the differential impacts of cyberattacks on EVs and ICE vehicles through comprehensive simulations, utilizing velocity standard deviation and spacing standard deviation as key performance metrics. Based on observed collision potentials, we classify the proposed cyberattacks into three distinct risk categories, providing a framework for risk assessment and mitigation strategies.

\end{itemize}
The rest of this paper is organized as follows.  
Section \ref{sec:METHODOLOGY} presents mathematical models for the six proposed communication-based attacks.  
Section \ref{sec:simulation results} describes the ring-road simulation setup, investigates the impacts of cyberattacks on EVs and ICE vehicles, and summarizes key findings.  
Finally, Section \ref{sec:conclusion} concludes this paper and outlines directions for future research.


\section{Methodology}\label{sec:METHODOLOGY}

We propose and mathematically formulate six distinct types of cyberattacks targeting ACC and communication systems. 
To systematically investigate how potential cyberattacks influence the behavior of AVs, this study employs a microscopic-level approach. 
The specific methodology for each attack is detailed in the subsequent subsections.
\subsection{A Brief Background and Car-following Model}
Various cyberattacks may target AVs, particularly those equipped with ACC systems in mixed traffic environments \cite{petit2014potential}. 
We begin by revisiting a mathematical model specifically designed to simulate potential attack scenarios in car-following dynamics \cite{li2024can, li2024detecting}. 
This model serves as an initial step to explore AV dynamics and assess the impact of cyberattacks on the overall traffic flow of ACC-enabled vehicles. Although these mathematical model scenarios may not directly represent real-world cyberattacks,
they provide valuable insights into the potential behavioral disruptions of ACC-equipped vehicles under adversarial conditions.

The microscopic car-following model serves as a fundamental framework for analyzing individual vehicle behavior \cite{sarafdissipation,li2024car, li2021classification}. This model characterizes a vehicle's acceleration as a function of its dynamic state and that of the preceding vehicle, typically incorporating parameters such as inter-vehicle gap and relative speed. The acceleration of a vehicle can be expressed by:
\begin{equation} 
a(t)= f(\boldsymbol{\theta}, v, s(t), \Delta v(t))
\end{equation}
where $a$ denotes the acceleration, $v$ is the speed of the following vehicle, and $s$ signifies the spacing between consecutive vehicles. The entity $\Delta v = v_l-v$ represents the relative speed, defined as the speed difference between the following vehicle $v$ and the preceding vehicle $v_l$, and $\boldsymbol{\theta}$ encapsulates model-specific parameters with time. Unless otherwise stated, the time index $t$ is omitted for simplicity whenever convenient.

\subsection{Discretized Packet-Dropping Attack}
AVs rely extensively on high-speed data transmission and reliable information exchange to enable timely and accurate driving decisions. Consequently, minimal communication delays or data loss can impair system responsiveness and substantially increase safety risks.
PDAs, frequently observed in ACC-equipped vehicles, primarily compromise V2V communication by selectively discarding critical messages containing positional and velocity information. This intentional disruption degrades inter-vehicle coordination and poses a significant threat to operational safety \cite{kim2021cybersecurity,dong2020impact}.

Recent studies by \cite{li2024can,hassan2020intelligent} have offered more comprehensive analyses of PDAs, demonstrating that these attacks indirectly compromise the reception of sensor data by inducing packet loss during network transmission. The discrepancy between received and actual data can cause erroneous assessments of road conditions, and may lead to accidents that endanger human safety in severe cases.

Therefore, investigating variants of PDAs is crucial for comprehensive cybersecurity assessment. Inspired by \cite{li2022detecting}, where communication delay is characterized as a function of time step $t$, we propose a PDA-based variant that discretizes the temporal component. We designate this attack as the Discretized Packet-Dropping Attack (DPDA), which causes the vehicle's motion to exhibit discrete, non-continuous behavior. The mathematical formulation for DPDA is expressed as:
\begin{multline}
    a(t)=(1-\delta) \cdot f(\boldsymbol{\theta}, v, s, \Delta v)+\\ \delta \cdot f\left(\boldsymbol{\theta}, v(t),  s(\lfloor t -m\rfloor), \Delta v(\lfloor t-m \rfloor)\right)
\end{multline}

\noindent where $\delta$ represents the attack indicator, with $\delta = 1$ when under attack and $\delta = 0$ otherwise. The acceleration $a(t) \in [\xi, \rho]$, where $\xi$ and $\rho$ denote the lower and upper physical bounds of vehicle acceleration, respectively. The variable $t$ represents the current time instance for normal information reception, while $\lfloor t - m \rfloor$ introduces a discretized delay component, with $m$ serving as the delay parameter. The floor function $\lfloor \cdot \rfloor$ ensures discrete time steps, creating abrupt changes in the perceived spacing $s$ and relative velocity $\Delta v$. This discretized representation more accurately captures the discontinuous velocity changes induced by cyberattacks, thereby enhancing the detectability and analysis of such malicious events.

\subsection{Phantom Attack}

During AV operation, sensors continuously acquire information through environmental perception mechanisms \cite{el2020cybersecurity}. The necessity to process large volumes of real-time information introduces vulnerabilities within the data transmission pipeline, which may be exploited by malicious actors to compromise system integrity.
To address this threat vector, we propose a novel category of cyberattack, referred to as the Phantom Attack (PA). Attackers inject falsified data into V2V communication channels, causing targeted AVs to erroneously receive and process information intended for other vehicles within the network.

In the adopted leader-follower vehicle dynamics model, each following vehicle receives information exclusively from its immediate predecessor \cite{wang2020modeling}. This information flow topology can be represented by an adjacency matrix $\bm{\mathrm{M}}  \in {0,1}^{N \times N}$, where $N$ denotes the total number of vehicles in the platoon. An element $\bm{\mathrm{M}}(i,j) = 1$ indicates an active information link from vehicle $j$ to vehicle $i$, while $\bm{\mathrm{M}}(i,j) = 0$ indicates no direct communication. The indices $i$ and $j$ correspond to the sequential positions of vehicles within the platoon and $i=1$ designates the lead vehicle:

\begin{equation*}
   \bm{\mathrm{M}}=
\begin{bmatrix}
    0 & 0 & \cdots & 0 & \cdots & 0 & 1 \\
    1 & 0 & \cdots & 0 & \cdots & 0 & 0 \\
    0 & 1 & \cdots & 0 & \cdots & 0 & 0 \\
    0 & 0 & \cdots & 1 & \cdots & 0 & 0 \\
    \vdots & \vdots & \ddots & \vdots & \ddots & \vdots & \vdots \\
    0 & 0 & \cdots & 0 & \cdots & 1 & 0 \\
    0 & 0 & \cdots & 0 & \cdots & 0 & 1 \\
\end{bmatrix}_{N\times N}
\end{equation*}

PA is designed to manipulate the vehicle communication structure by inducing sensors to receive and process falsified information. In particular, when a vehicle receives falsified inputs for relative velocity $\Delta v$ and inter-vehicle spacing $s$, its control system may generate erroneous acceleration commands, potentially leading to unsafe driving behaviors and platoon instability. Based on this communication topology analysis, we formulate the mathematical model for the Phantom Attack where vehicle $i$ erroneously receives information intended for vehicle $j$:
\begin{multline}
     a_i(t) = (1-\delta) \cdot f(\boldsymbol{\theta}, v, s_i, v_{i-1} - v_{i}) +\\  \delta \cdot f(\boldsymbol{\theta}, v,  s_j, v_{j-1} - v_{j})
\end{multline}  
where $\delta$ represents the attack indicator ($\delta = 1$ during attack, $\delta = 0$ otherwise). The term $a_i(t)$ denotes the acceleration of the $i$-th vehicle at time step $t$, while $s_i$ and $s_j$ represent the inter-vehicle spacing for vehicles $i$ and $j$ with respect to their immediate predecessors, respectively. The velocity terms $v_{i-1} - v_i$ and $v_{j-1} - v_j$ represent the relative velocities between consecutive vehicle pairs. This formulation captures how the attack causes vehicle $i$ to compute its acceleration based on the traffic conditions experienced by vehicle $j$, potentially resulting in dangerous driving behaviors when vehicles $i$ and $j$ experience significantly different traffic scenarios.

\subsection{Fixed-Speed Attack}
Cyberattacks targeting engine control systems and vehicle velocity have received considerable attention in recent cybersecurity research \cite{stabili2022exploring}. We propose a novel attack vector, referred to as the Fixed-Speed Attack (FA). In contrast to conventional attacks that manipulate sensor inputs or intercept real-time data \cite{li2024can}, FA directly compromises the ACC system, forcing the targeted vehicle to maintain a constant velocity regardless of dynamic traffic conditions.

In this attack scenario, the vehicle’s velocity is locked at the value observed at the moment of attack initiation and maintained at this fixed speed for the duration of the attack. By targeting the velocity control mechanism of ACC, FA prevents the vehicle from executing necessary speed adjustments based on real-time traffic conditions. Consequently, the compromised vehicle loses its capacity to respond appropriately to its surroundings, potentially causing operational hazards, traffic flow disturbances, and even catastrophic collisions. The mathematical formulation for the FA is expressed as:
\begin{equation}
    a(t)=(1-\delta) \cdot  f(\boldsymbol{\theta}, v, s, \Delta v )   +\delta \cdot f(\boldsymbol{\theta}, v(\tilde{t}),  s, \Delta v  ) 
\end{equation}
where $\delta$ denotes the attack indicator ($\delta = 1$ during attack, $\delta = 0$ otherwise), and $\tilde{t}$ represents the attack initiation time. Under this attack, the vehicle's velocity parameter in the ACC control function is fixed at $v(\tilde{t})$, which is the velocity at the moment of attack initiation. 
Although the actual spacing $s$ and relative velocity $\Delta v$ continue to vary with traffic conditions, the ACC system computes acceleration based on the compromised velocity input $v(\tilde{t})$ instead of the actual current velocity.

This manipulation effectively corrupts the ACC's control logic, causing it to generate inappropriate acceleration commands that attempt to maintain the vehicle at a constant speed, thereby overriding the system's adaptive behavior and creating potentially hazardous driving conditions.

\subsection{Blinding Attack}
AVs rely on sophisticated sensor arrays and communication systems for safe operation \cite{el2020cybersecurity}. These critical components, while essential for vehicle autonomy, present attractive targets for malicious actors seeking to compromise vehicle safety. In this context, we introduce a novel cyberattack vector termed the Blinding Attack (BA), which systematically disrupts the vehicle's environmental perception capabilities.
The BA operates by simultaneously compromising multiple sensing modalities, including V2V communication channels and V2I infrastructure links. By corrupting these perception mechanisms, BA fundamentally impairs the ACC system's ability to accurately detect and respond to surrounding traffic conditions. 
The name of the attack originates from its ability to create selective blind spots within the vehicle’s perception system, causing the system to overlook critical obstacles, particularly intermediate vehicles in the forward path. This oversight may lead to severe safety incidents with potentially catastrophic consequences.

In the standard leader-follower vehicle model \cite{wang2020modeling}, each vehicle responds to its immediate predecessor. However, the targeted vehicle's perception system is manipulated to disregard intermediate vehicles under BA, causing it to respond instead to a more distant vehicle as if no intervening traffic exists. This perceptual distortion creates a hazardous scenario in which the attacked vehicle makes control decisions based on incomplete environmental information. The acceleration dynamics of a vehicle subjected to the BA are formulated as:
\begin{equation}
    a(t)=(1-\delta ) \cdot  f(\boldsymbol{\theta}, v, s_i,  v_{i-1}-v_i)+ \delta \cdot f(\boldsymbol{\theta}, v, \tilde{s} , v_j-v_i)
\end{equation}
where $\delta$ denotes the attack indicator ($\delta = 1$ during attack, $\delta = 0$ otherwise). The term $s_i$ represents the spacing between vehicles $i$ and $i-1$ under normal conditions. The relative velocity term shifts from $v_{i-1} - v_i$ (with respect to the immediate predecessor) to $v_j - v_i$ (with respect to the distant vehicle $j$) under BA.
Additionally, the perceived spacing $\tilde{s}$ under BA denotes the cumulative distance from vehicle $i$ to $j$, aggregating all intermediate inter-vehicle gaps. 
It should be noted that this perceived spacing should not exceed a realistic threshold since excessively large values may deviate from plausible real-world conditions. Hence, the perceived spacing between vehicle $i$ and $j$ is formally defined as:
\begin{equation}\label{eq:BA}
  \tilde{s} = \left\{
  \begin{array}{ll}
    \sum_{z=0}^{p}s_{(i-z)\bmod N}  & \mbox{ if } \sum_{z=0}^{p}s_{(i-z)\bmod N}  < \varphi    \\
    \varphi & \mbox{ if }  \sum_{z=0}^{p}s_{(i-z)\bmod N}  \geq \varphi 
  \end{array}
  \right.
\end{equation}
where $p=(i-j-1) \bmod N$ represents the number of blinded vehicles under BA, and $N$ denotes the total number of vehicles in the platoon. The entity $\varphi$ is a user-defined threshold for maximum  perceived spacing under BA.

This attack formulation captures a particularly insidious scenario in which the attacked vehicle perceives an artificially inflated gap ahead and responds to the velocity of a distant vehicle rather than that of its immediate predecessor, potentially executing inappropriate acceleration maneuvers.
Such behavior dramatically elevates the risk of rear-end collisions with the undetected intermediate vehicles, as the ACC system operates under the false assumption of a clear road ahead. The severity of this attack lies in its ability to exploit the fundamental trust that ACC systems place in their sensory inputs, converting a safety-critical system into a potential hazard.

\subsection{Angular Velocity Attack}
Traditional cyberattacks on AVs primarily target linear motion parameters such as velocity and acceleration. In contrast, we introduce a novel attack vector that exploits the relationship between angular and linear motion, termed the Angular Velocity Attack (AVA). This attack uniquely targets the IMU, corrupting orientation data to induce unintended rotational components in the vehicle's motion while maintaining forward velocity. Through IMU manipulation, AVA indirectly compromises V2V communication by causing transmitted positional and directional data to deviate from the vehicle's actual trajectory.

The AVA simulates a scenario in which malicious actors inject false angular velocity readings into the vehicle's control system, causing the ACC to compute longitudinal acceleration based on corrupted velocity projections. Although the vehicle maintains its physical forward motion, the attack introduces a virtual rotation component that affects the velocity value used in ACC calculations. This discrepancy between actual and perceived motion can cause significant trajectory deviations, potentially resulting in lane departures or catastrophic collisions.

For analytical tractability, this initial investigation constrains AVA to single-lane scenarios, where lateral motion is restricted but longitudinal dynamics remain affected by the angular component. Extension to multi-lane environments, where the full impact of angular deviations becomes apparent, represents a valuable direction for future research.
Drawing inspiration from GPS spoofing research \cite{ying2023gps}, we formulate the AVA by incorporating heading angle deviations into the ACC control function by:
\begin{equation}
    a(t)=(1-\delta) \cdot f(\boldsymbol{\theta}, v,  s , \Delta v )  +\delta \cdot f(\boldsymbol{\theta}, v \cdot G(\phi), s, \Delta v )
    \label{AVA-eq}
\end{equation}

\noindent where $\delta$ represents the attack indicator ($\delta = 1$ during attack, $\delta = 0$ otherwise), and $\phi$ denotes the induced heading angle deviation. The function $G(\phi)$ represents a trigonometric transformation, specifically either $\sin(\phi)$ or $\cos(\phi)$, which modulates the effective velocity input to the ACC system. Under this attack, the vehicle's velocity component used for acceleration calculations becomes $v \cdot G(\phi)$ rather than the true velocity $v$.

This formulation captures how angular velocity corruption manifests in longitudinal dynamics: when $G(\phi) = \cos(\phi)$, the attack simulates a reduction in the forward velocity component as if the vehicle were traveling at an angle to its intended path. Conversely, when $G(\phi) = \sin(\phi)$, the attack models the projection of velocity onto a perpendicular axis, creating more severe disruptions to the ACC calculations. 
The corrupted velocity values result in erroneous acceleration commands, which can lead to significant speed fluctuations and degradation of vehicle stability, especially when the induced angle $\phi$ becomes substantial.

\subsection{Mixed Attack}

AVs operating in real-world environments face numerous disturbances, including electromagnetic interference, signal attenuation, and adverse environmental conditions \cite{zhang2023perception}. These challenges necessitate the investigation of sophisticated attack scenarios that more accurately represent realistic threat landscapes. To address this need, we propose a Mixed Attack (MA) framework that combines multiple attack vectors to create compound cyber threats against V2V communication systems.

In this study, we specifically implement MA by integrating the DPDA and PA mechanisms. This hybrid strategy disrupts cooperative vehicle behavior through simultaneous manipulation of both temporal and informational aspects of V2V communication. The DPDA component introduces temporal discontinuity by creating discrete time delays in information processing, while the PA component injects falsified data from incorrect vehicle sources. These attack mechanisms target distinct yet complementary vulnerabilities: DPDA exploits timing-based weaknesses in the communication protocol, whereas PA exploits trust assumptions in the information source verification process.

The selection of DPDA and PA for this MA implementation is motivated by several factors. Firstly, both attacks operate within the same V2V communication layer, enabling seamless integration without requiring access to multiple system components. Secondly, their combined effect creates a synergistic disruption pattern that is significantly more challenging to detect and mitigate than either attack in isolation. Thirdly, this combination offers an optimal balance between attack sophistication and implementation complexity, making it particularly suitable for evaluating system resilience under compound adversarial conditions.
The acceleration dynamics of ACC-equipped vehicle $i$ subjected to the MA are formulated as:
\begin{multline}
      a_i(t)=(1-\delta) \cdot f(\boldsymbol{\theta}, v, s_i, v_{i-1}-v_i ) +\\ \delta \cdot f(\boldsymbol{\theta}, v(t),  s_j(\lfloor t -m\rfloor),  v_{j-1}(\lfloor t-m \rfloor)-v_j(\lfloor t-m \rfloor))
\end{multline}
where $\delta$ represents the attack indicator ($\delta = 1$ during attack, $\delta = 0$ otherwise). The spacing terms $s_i$ and $s_j$ denote the inter-vehicle distances for vehicles $i$ and $j$ relative to their respective predecessors. The discretized time expression $\lfloor t - m \rfloor$ implements the DPDA component by introducing temporal quantization with delay parameter $m$, while the substitution of vehicle $j$'s information for vehicle $i$'s expected data implements the PA component. This formulation captures the compound effect whereby vehicle $i$ receives outdated information from an incorrect source, potentially causing severe control instabilities when vehicles $i$ and $j$ experience divergent traffic conditions.

While this study focuses on the DPDA-PA combination, the MA framework is extensible to incorporate additional attack vectors. Future research may explore alternative attack combinations to comprehensively evaluate the resilience of ACC systems against multi-dimensional cyber threats.

\subsection{Summary}

The preceding sections have presented comprehensive mathematical formulations for six novel cyberattack vectors targeting AVs, elucidating their fundamental mechanisms and dynamic characteristics. To facilitate systematic comparison and risk assessment, this section synthesizes our findings from an adversarial perspective, analyzing the environmental conditions that maximize each attack's effectiveness and evaluating their potential consequences on traffic safety.

Table \ref{tab:summary} provides a structured overview of the proposed cyberattacks, encompassing their operational descriptions, vulnerability-inducing scenarios, and projected impacts on vehicular systems. The ‘Potential Real-world Scenarios’ column identifies specific traffic environments and conditions where each attack vector exhibits maximum exploitability, considering factors such as infrastructure limitations, environmental interference, and traffic density. Our analysis reveals a concerning pattern: while the immediate manifestations vary, all examined cyberattacks ultimately compromise traffic flow stability. Under high-density traffic conditions, these disruptions can rapidly cascade into severe safety incidents, including multi-vehicle collisions and lane departure accidents, thereby presenting substantial threats to both infrastructure integrity and human safety.

The relative severity and deployment feasibility of each attack vector will be quantitatively evaluated through comprehensive simulation studies presented in the subsequent sections. These empirical assessments will provide critical insights for developing targeted countermeasures and enhancing the resilience of vehicle systems.

\begin{table*}[htbp]
    \centering
    \footnotesize
    \renewcommand{\arraystretch}{1.2}
    \caption{\\Summary of different cyberattack types.}
    \resizebox{\textwidth}{!}{
    \begin{tabular}{>{\centering\arraybackslash}m{1.5cm} >{\centering\arraybackslash}m{4.5cm} >{\centering\arraybackslash}m{3.5cm} >{\centering\arraybackslash}m{3.5cm}}
    \hline
    \textbf{Attack} & \textbf{Description} & \textbf{Potential Real-world Scenarios}& \textbf{Possible Consequences}  \\
    \hline
    
    DPDA & Introduces discretized temporal delays in information reception through V2V communication interference. & Infrastructure-limited areas with degraded signal quality (e.g., mountainous terrain, tunnel systems). & High collision probability when attacked vehicles operate in adjacent positions.\\
    \hline
    PA & Causes vehicles to receive falsified information from incorrect sources via compromised V2V communication. & High-density traffic environments (e.g., multi-lane highways, urban expressways). & Severe traffic flow disruption and coordination failure. \\ 
    \hline
    FA & Forces vehicle to maintain constant velocity at attack initiation moment by overriding ACC control. & Confined roadways with limited maneuvering space (e.g., tunnels, single-lane rural roads). & Critical collision risk in adjacent vehicle configurations. \\
    \hline
    BA & Manipulates vehicle perception causing incorrect detection of preceding vehicles through V2V and V2I compromise. & Adverse visibility conditions (e.g., dense fog, heavy precipitation, dust storms). & Direct vehicle collision with undetected obstacles. \\
    \hline
    AVA & Introduces angular velocity components affecting longitudinal dynamics via IMU and V2V attacks. & Geometrically complex roads (e.g., serpentine mountain passes, curved elevated roads). & Platoon stability disruption and formation breakdown. \\
    \hline
    MA & Synergistic combination of DPDA and PA simultaneously targeting V2V communication channels. & Topologically complex traffic scenarios (e.g., multi-level interchanges, roundabouts). & Compounded traffic flow instability and unpredictable behaviors. \\
    \hline
    \end{tabular}
    }
    \label{tab:summary}
\end{table*}

\begin{table*}[b]
\centering
\caption{\\Simulation scenarios and vehicle configurations with unique indices.}
\begin{tabular}{>{\centering\arraybackslash}p{3cm}
>{\centering\arraybackslash}p{3cm}
>{\centering\arraybackslash}p{3cm}
>{\centering\arraybackslash}p{7.4cm}}
\hline
\textbf{Scenario} & \textbf{ACC Vehicles} & \textbf{Attacked Vehicles} & \textbf{Description} \\
\hline
I & [1,6] & [1] & Single vehicle attack \\
\hline
II & [1,3,5,7] & [1,5] & Two non-adjacent attacks \\
\hline
III & [1,3,5,7,9] & [1,5,9] & Three non-adjacent attacks \\
\hline
IV & [1,3,5,6,7] & [1,5,6] & Adjacent attacks (vehicles 5 and 6) \\
\hline
\end{tabular}
\label{Simulated scenarios and vehicle numbering}
\end{table*}

\section{Simulation and Experiment}\label{sec:simulation results}
To investigate the impact of the proposed cyberattacks on vehicle dynamics, we conduct comprehensive simulations involving both EVs and ICE vehicles equipped with ACC systems. This section presents the experimental framework, simulation methodology, and evaluation metrics employed in our analysis.

\subsection{Experimental Setup}

\begin{figure}[htbp]
\centering
\subfloat[Experimental circular road layout.\label{fig:circular-road}]{\includegraphics[width=0.60\textwidth]{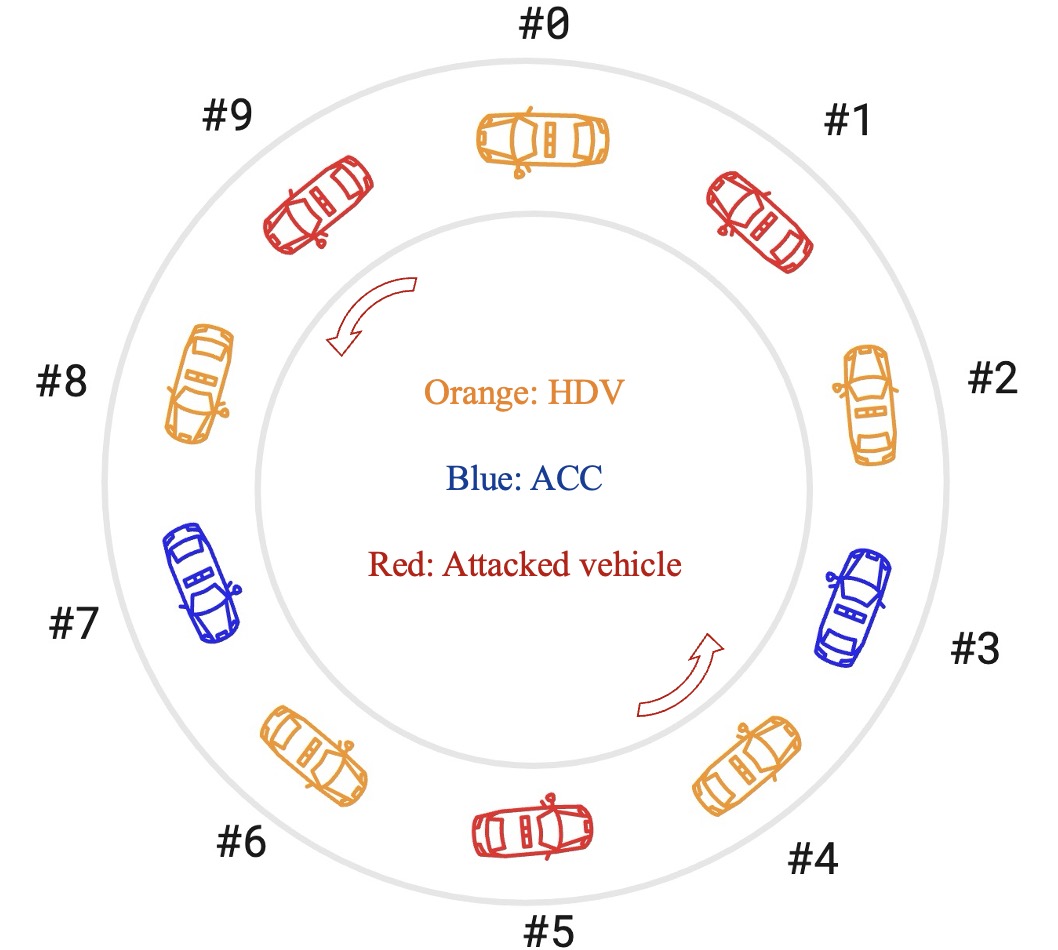}}

\vspace{0em}
\subfloat[Attack configuration for Scenario III.\label{fig:attack-details}]{
    \includegraphics[width=0.60\textwidth]{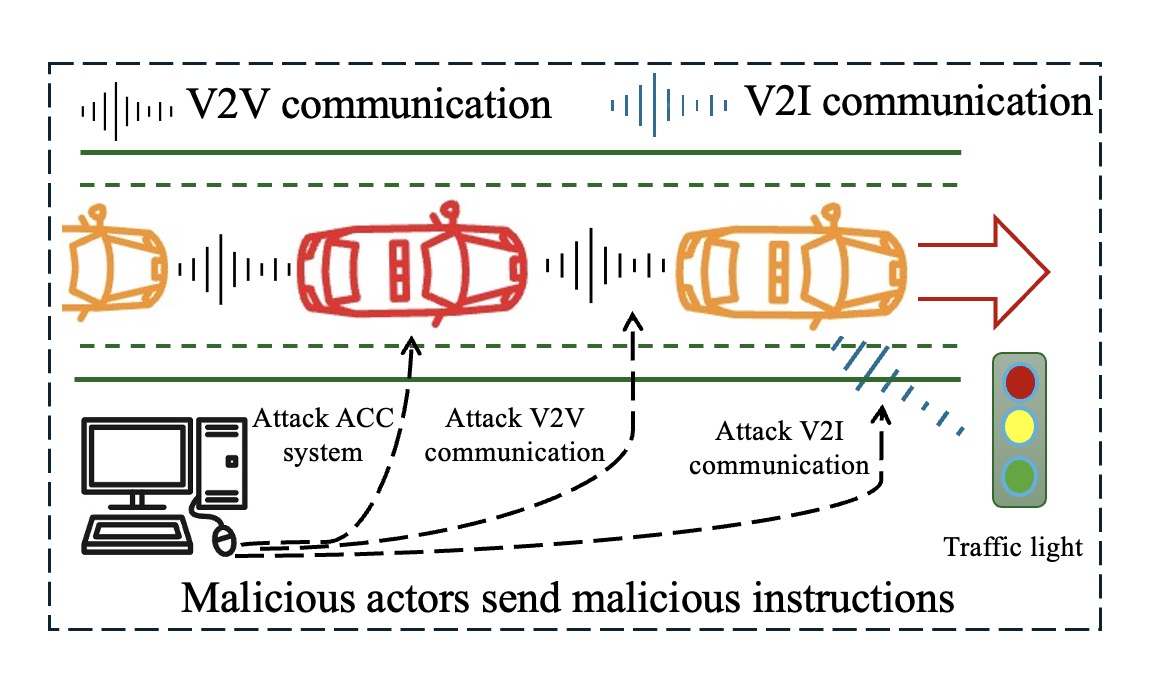}
}
\caption{Initial configuration of simulated Scenario III where three non-adjacent vehicles in red are subjected to targeted cyberattacks. The direction of traffic flow is counterclockwise.}
\label{fig:scenario-layout}
\end{figure}

The simulation environment consists of a 300-meter circular roadway, as illustrated in Fig. \ref{fig:circular-road}. This ring-road configuration is widely adopted in traffic flow studies to effectively simulate continuous leader-follower dynamics without boundary effects \cite{sugiyama2008traffic, giammarino2020traffic}. 
Ten vehicles are uniformly distributed along the circuit with equal inter-vehicle spacing of $25$ meters, and all vehicles are initially stationary.

The vehicle fleet comprises a combination of Human-Driven Vehicles (HDVs) and ACC-equipped vehicles \cite{wang2023novel}, with the latter category including both ICE and EV platforms. 
To comprehensively evaluate the impact of the proposed cyberattack types, we design four distinct simulation scenarios in Table \ref{Simulated scenarios and vehicle numbering}. These scenarios systematically vary ACC MPRs and attack target configurations, ranging from single to multiple compromised vehicles arranged in both adjacent and non-adjacent topologies, as illustrated in Fig. \ref{fig:attack-details}.
The simulation is conducted over a 120-second timeline, which is partitioned into three distinct phases: pre-attack (30 to 60 seconds), during-attack (60 to 90 seconds), and post-attack (90 to 120 seconds).
These phases are visually delineated by shaded regions in the subsequent figures. Vehicle car-following behavior is modeled using the Intelligent Driver Model (IDM), which has been widely adopted to effectively capture realistic human driving dynamics \cite{treiber2000congested,talebpour2016influence}.
The kinematic state of the $k$-th vehicle evolves according to the following dynamic formulation:
\begin{equation}
\begin{bmatrix}
s_k \\
v_k
\end{bmatrix}
=
\begin{bmatrix}
s_k \\
v_k
\end{bmatrix}_t
+
\begin{bmatrix}
v_{k-1} - v_k \\
f_k
\end{bmatrix}_t \Delta t
\label{kinematic equations}
\end{equation}
where $1 < k \leq 10$ denotes the vehicle index, and the acceleration $f_k$ can be computed by Eq. (\ref{IDM}). The simulation employs a temporal resolution of $\Delta t = 0.033$ seconds, corresponding to an update frequency of 30 Hz.
The IDM acceleration function is formulated as:
\begin{equation}
f(\boldsymbol{\theta},s,v,\Delta v)= \alpha \left[1-\left(\frac{v}{v_d}\right)^\kappa-\left(\frac{\hat{s}(v,\Delta v)}{s}\right)^2\right] \label{IDM}
\end{equation}
where the desired spacing $\hat{s}(v,\Delta v)$ is given by:
\begin{equation}
\hat{s}(v,\Delta v)= \eta + \tau v + \frac{v\Delta v}{2\sqrt{\alpha\beta}}
\end{equation}

The IDM parameter vector $\boldsymbol{\theta} = [\alpha, \beta, \kappa, \eta, \tau, v_d]^\top$ encapsulates vehicle-specific behavioral characteristics, where $\alpha$ represents maximum acceleration, $\beta$ denotes comfortable deceleration, $\kappa$ is the acceleration exponent, $\eta$ specifies minimum spacing, $\tau$ defines the desired time gap, and $v_d$ indicates the desired velocity.
Table \ref{tab:idm_parameters} presents the calibrated IDM parameters for each vehicle type. Particularly, ICE-ACC parameters are derived from \cite{de2021calibrating} and validated in \cite{gunter2020commercially}. EV-ACC parameters adopt the medium-gap configuration from \cite{zare2023modeling} to accurately represent typical EV driving characteristics. HDV parameters are sourced from \cite{kesting2008calibrating}, reflecting empirically observed human driving behavior.

\begin{table}[htbp]
\centering
\caption{ \\ IDM parameter sets for different vehicle types.}
\label{tab:idm_parameters}
\begin{tabular}{ccccccc}
\hline
\multirow{3}{*}{\textbf{Vehicle Type}} & \multicolumn{6}{c}{\textbf{IDM Parameters}}                                \\ \cline{2-7} 
& $\alpha$& $\beta$& $\kappa$& $\eta$& $\tau$& $v_d$\\
& $(m/s^2)$& $(m/s^2)$& -& $(m)$& $(s)$& $(m/s)$\\ 
\hline
EV-ACC                        &         2.01&        8.97&       4.02&          2.02&           1.63&           33.34\\
ICE-ACC                       &         0.60&        5.20&       15.50&          6.30&           2.20&           44.11\\
HDV                           &         1.06&        2.00&       4.00&          3.40&           1.26&           30.00\\ \hline
\end{tabular}

\end{table}

\renewcommand{\arraystretch}{1.2}
\captionsetup[table]{labelsep=space, justification=centering, textfont=sc}
\begin{table*}[h]
\centering
\caption{\\ \MakeUppercase{Baseline performance of EVs and ICE vehicles across different simulation scenarios in the absence of cyberattacks. Units: $V_{avg}$ (m/s), $\overline{VSD}$ (m/s), $\overline{SSD}$ (m), $THW$ (s).} }

\resizebox{\textwidth}{!}{
\begin{tabular}{clp{1.9cm}p{1.9cm}p{1.9cm}p{1.9cm}p{1.9cm}p{1.9cm}}
\hline
\multirow{2}{*}{\textbf{Scenario}} & \multirow{2}{*}{\textbf{Metric}} & \multicolumn{3}{c}{\textbf{EV}} & \multicolumn{3}{c}{\textbf{ICE Vehicle}} \\
\cmidrule(lr){3-5} \cmidrule(lr){6-8}
\multicolumn{2}{l}{} & \textbf{Pre} & \textbf{During} & \textbf{Post} & \textbf{Pre} & \textbf{During} & \textbf{Post} \\
\hline
\multirow{4}{*}{I} & \textbf{$V_{avg}$} & 15.59 & 15.74 & 15.74 & 13.78 & 14.19 & 14.19 \\
& $\overline{VSD}$ & 0.20 & 0.06 & 0.08 & 0.38 & 0.27 & 0.12 \\
& $\overline{SSD}$ & 0.34 & 0.15 & 0.18 & 0.95 & 0.74 & 0.38 \\
& $THW$ & 1.60 & 1.59 & 1.59 & 1.83 & 1.78 & 1.77 \\
\hline
\multirow{4}{*}{II} & \textbf{$V_{avg}$} & 15.12 & 15.26 & 15.21 & 11.89 & 12.27 & 12.28 \\
& $\overline{VSD}$ & 0.22 & 0.11 & 0.10 & 0.75 & 0.63 & 0.80 \\
& $\overline{SSD}$ & 0.41 & 0.30 & 0.27 & 2.21 & 2.33 & 2.84 \\
& $THW$ & 1.66 & 1.64 & 1.64 & 2.13 & 2.04 & 2.01 \\
\hline
\multirow{4}{*}{III} & \textbf{$V_{avg}$} & 14.89 & 15.01 & 14.98 & 11.34 & 11.49 & 10.93 \\
& $\overline{VSD}$ & 0.23 & 0.22 & 0.23 & 0.57 & 0.81 & 1.04 \\
& $\overline{SSD}$ & 0.65 & 0.59 & 0.63 & 2.47 & 3.12 & 4.21 \\
& $THW$ & 1.68 & 1.67 & 1.67 & 2.23 & 2.15 & 2.14 \\
\hline
\multirow{4}{*}{IV} & \textbf{$V_{avg}$} & 14.88 & 15.00 & 14.94 & 11.08 & 10.97 & 11.43 \\
& $\overline{VSD}$ & 0.47 & 0.34 & 0.34 & 1.10 & 1.19 & 1.64 \\
& $\overline{SSD}$ & 1.15 & 1.10 & 1.06 & 3.69 & 4.54 & 5.78 \\
& $THW$ & 1.68 & 1.67 & 1.67 & 2.30 & 2.26 & 2.14 \\
\hline
\end{tabular}
}
\label{no attack scenarios}
\end{table*}

Quantitative assessment of vehicle interactions under cyberattack conditions requires behaviorally relevant metrics. Time Headway (THW), a key indicator of longitudinal safety and traffic flow stability, is widely used in autonomous driving control and traffic risk assessment \cite{khansari2020study}.
Recent studies under steady-state conditions have shown that THW varies significantly with speed, driving environment, and individual driver characteristics \cite{parashar2025reassessing}, underscoring its importance in traffic behavior modeling and safety strategy design.
Under collision-free conditions, smaller THW values indicate shorter distances between vehicles, reflecting a more compact traffic flow. The THW is defined as follows:
\begin{equation} 
THW=  \frac{d}{v} 
\end{equation}
where $d$ represents the inter-vehicle distance and $v$ denotes the velocity of the following vehicle.

Vehicle speed stability constitutes a key indicator of operational safety and passenger comfort, typically evaluated through velocity standard deviation metrics \cite{hamzeie2017driver}. To comprehensively capture temporal variations in vehicle dynamics across different attack phases, we introduce two complementary evaluation metrics: Velocity Standard Deviation (VSD) and Spacing Standard Deviation (SSD).
The VSD for vehicle $i$ during a given time period $T$ can be computed by:
\begin{equation}
   VSD_i=\sqrt{\frac{1}{T} \sum_{t=1}^T(v_{i}-\bar v_i)^2}
\end{equation}
where $\bar{v}_i =(1/T) \sum_{t=1}^T v_{i,t}$ denotes the temporal average velocity, and  $v_{i,t}$ represents the velocity of vehicle $i$ at time step $t$. The fleet-wide average $VSD$ is computed as:
\begin{equation}
\overline{VSD} =\frac{1}{N} \sum_{i=1}^N VSD_i
\end{equation}
An elevated $\overline{VSD}$ value indicates greater velocity instability and erratic driving behavior, whereas a lower value suggests smoother and more stable vehicle operation.
Similarly, the $SSD$ for vehicle $i$ is formulated as:
\begin{equation}
   SSD_i=\sqrt{\frac{1}{T} \sum_{t=1}^T(s_{i}-\bar s_i)^2}
\end{equation}
where $\bar{s}_i = (1/T)\sum_{t=1}^T s_{i,t}$ denotes the temporal average spacing, and $s_{i,t}$ represents the spacing at time step $t$. The fleet-wide average $SSD$ is given by:
\begin{equation}
\overline{SSD} = \frac{1}{N} \sum_{i=1}^N SSD_i
\end{equation}
\noindent In addition, the overall Average Speed (denoted as $V_{avg}$) reflects the operational efficiency of the traffic system, and serves as a fundamental reference for comparing performance under different abnormal interventions by:
\begin{equation}
    V_{avg}= \frac{1}{N} \sum_{i=1}^N \bar v_i
\end{equation}
where $\bar v_i$ denotes the average speed of vehicle $i$.

A lower $\overline{SSD}$ value indicates stable and uniform vehicle spacing, while a higher value reflects increased variability potentially leading to traffic flow instability or congestion formation. 
Similarly, a smaller $V_{avg}$ value reflects lower overall vehicle speeds, whereas a larger $V_{avg}$ indicates that the vehicle fleet is moving at a generally higher speed.
A greater $\overline{VSD}$ suggests more pronounced speed variations during driving, while a reduced $\overline{VSD}$ indicates smoother and more stable vehicle motion.
These indicators provide an objective basis for quantifying the potential impact of cyberattacks on vehicle dynamics. 
Accordingly, assessing the impact of cyberattacks on speed and spacing fluctuations in EVs and ICE vehicles is critical for characterizing the dynamic response behaviors of different vehicle types under adversarial conditions.
The incorporation of $\overline{VSD}$ and $\overline{SSD}$ enables a more comprehensive evaluation and comparison of vehicle stability across various attack scenarios.

\subsection{Baseline Performance Analysis}

\begin{figure}[htbp]
\centering
\subfloat[EV spacing.\label{fig:ev-spacing-baseline}]{
    \includegraphics[width=0.45\columnwidth]{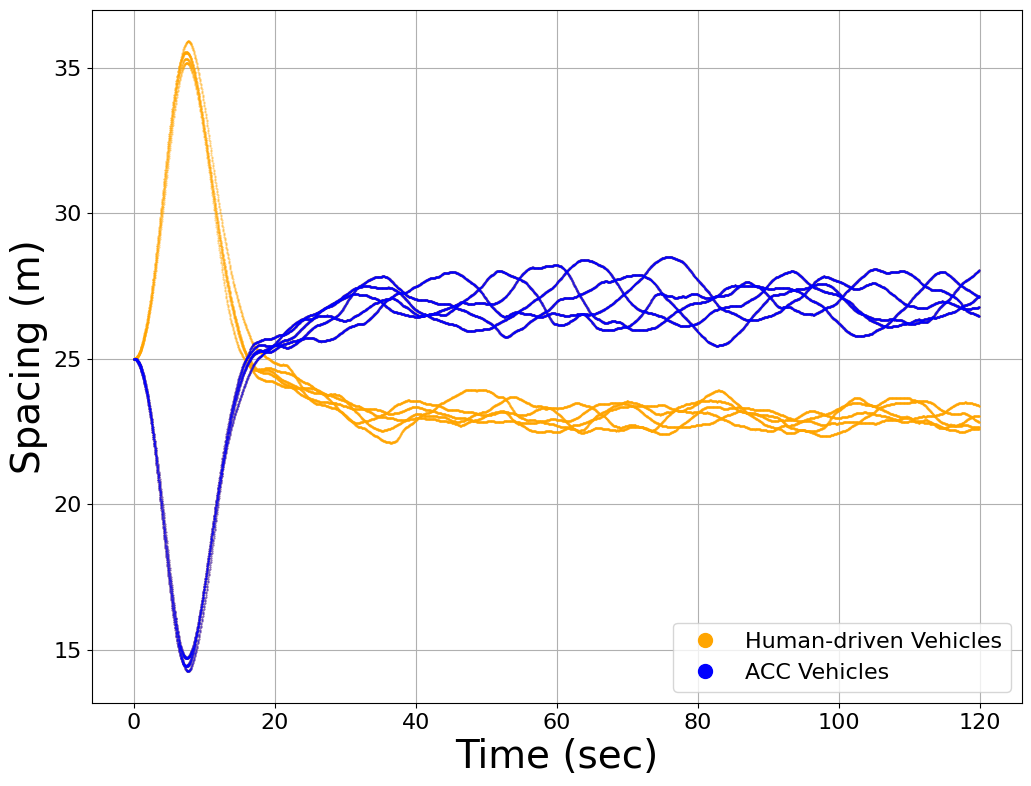}
}
\hfill
\subfloat[EV speed.\label{fig:ev-speed-baseline}]{
    \includegraphics[width=0.45\columnwidth]{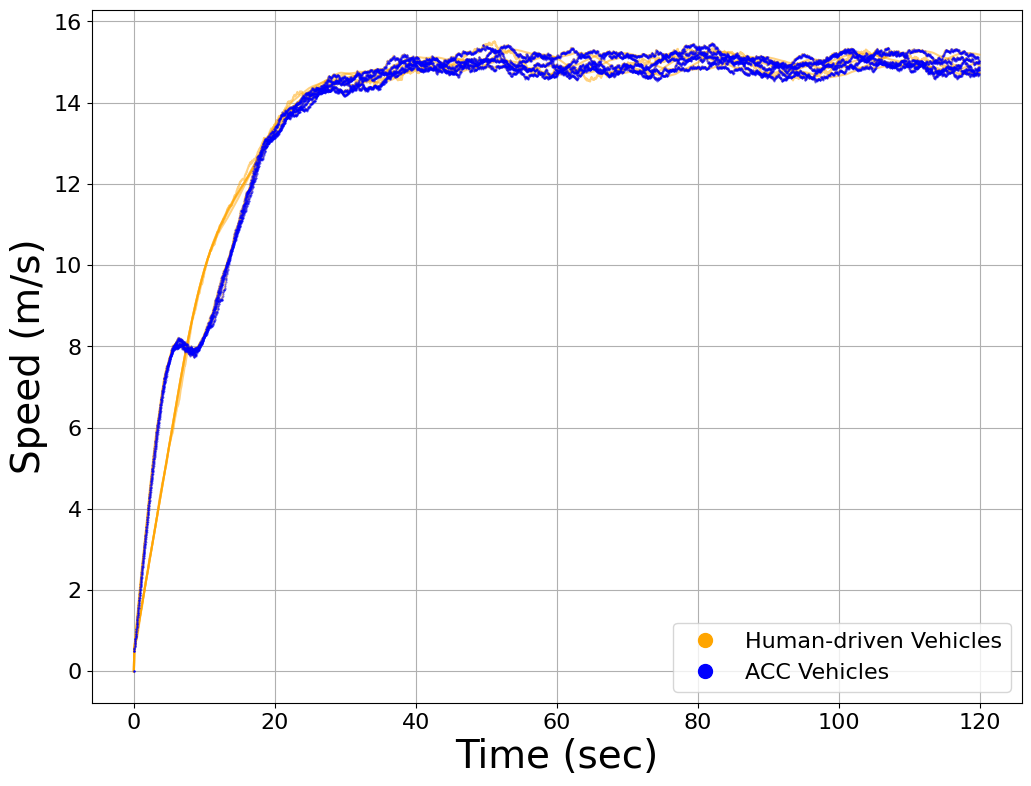}
}

\vspace{0em}  

\subfloat[ICE vehicle spacing.\label{fig:ice-spacing-baseline}]{
    \includegraphics[width=0.45\columnwidth]{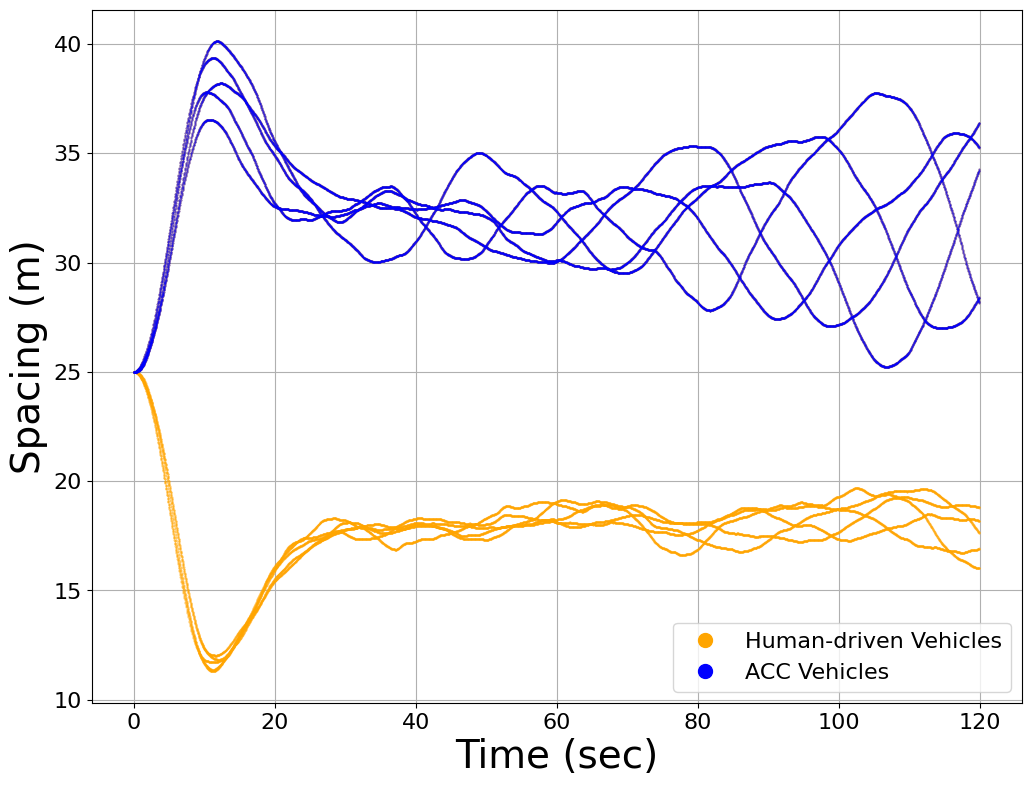}
}
\hfill
\subfloat[ICE vehicle speed.\label{fig:ice-speed-baseline}]{
    \includegraphics[width=0.45\columnwidth]{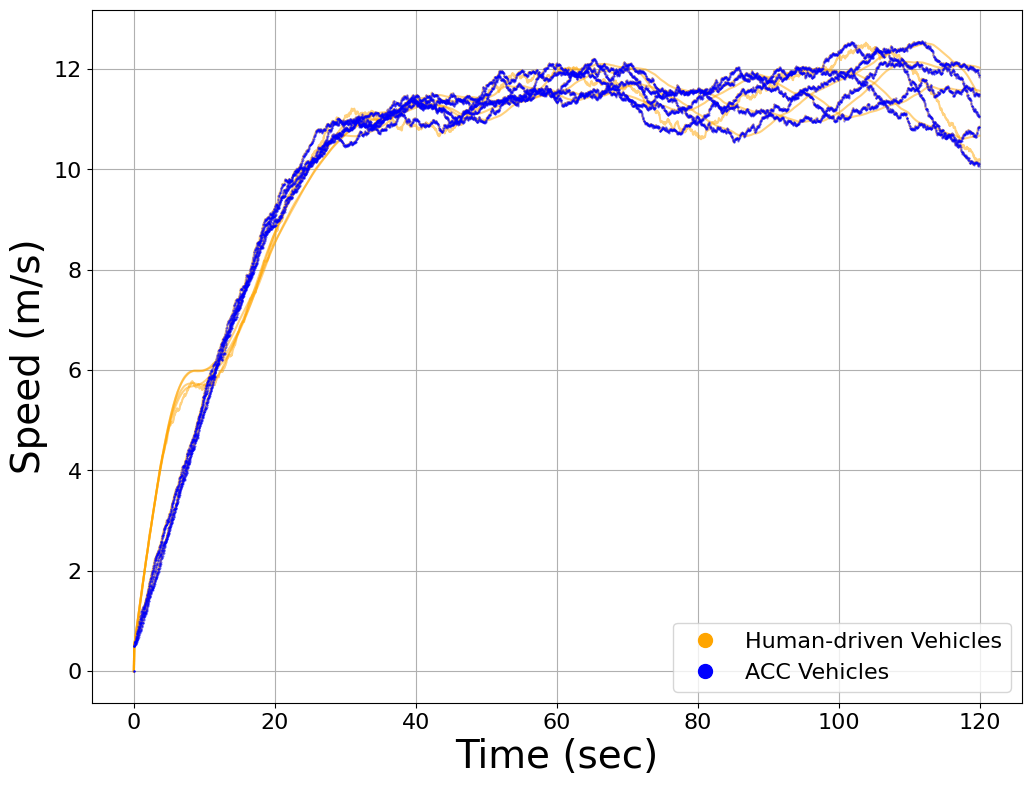}
}

\caption{Baseline spacing and velocity profiles for EVs and ICE vehicles in Scenario III under normal operating conditions (no cyberattacks).}
\label{fig:scenario3-baseline}
\end{figure}

To establish a comprehensive benchmark for evaluating the proposed cyberattacks, we first analyze vehicle dynamics under normal operating conditions across all simulation scenarios in Table \ref{Simulated scenarios and vehicle numbering}. 
For brevity, this section presents the baseline dynamics for Scenario III in Fig. \ref{fig:scenario3-baseline}, which incorporates five ACC-equipped vehicles and three designated target positions configured in a non-adjacent layout, as depicted in Fig. \ref{fig:circular-road}. No cyberattacks are applied in this scenario.
The comparative analysis reveals distinct behavioral patterns between EVs and ICE vehicles. Specifically, EVs consistently maintain shorter following distances (Fig. \ref{fig:ev-spacing-baseline} vs. \ref{fig:ice-spacing-baseline}) while achieving higher operational velocities (Fig. \ref{fig:ev-speed-baseline} vs. \ref{fig:ice-speed-baseline}), indicating superior dynamic performance and platoon cohesion.

Table \ref{no attack scenarios} presents the baseline performance across different simulation phases, evaluated using metrics including $V_{avg}$, $\overline{VSD} $, $\overline{SSD} $, and $THW$. 
Results indicate that Scenario I experiences the least impact, whereas Scenario IV exhibits the most severe disruptions, primarily due to the adjacency of attacked vehicles. 
This is attributable to Scenario I exhibiting the lowest $\overline{VSD} $ and $\overline{SSD} $ values, signifying minimal initial disturbance to traffic flow. In contrast, Scenario IV records the highest $\overline{VSD} $ and $\overline{SSD} $ values, indicating the most pronounced initial impact on traffic dynamics.
EVs exhibit lower $THW$ and higher $V_{avg}$ compared to ICE vehicles, indicating that EVs operate at higher speeds and maintain a more compact platoon formation.

\subsection{Discretized Packet-Dropping Attack Analysis}

To evaluate the impact of DPDA on vehicle dynamics, simulations were conducted across three discretized delay parameters: 6, 8, and 9 seconds, respectively. Table \ref{Combinations of Delay Times and Scenarios} summarizes the collision outcomes across all scenarios and delay configurations.

Results highlight a significant vulnerability in Scenario IV, as it is the only case in which collisions occurred for both EVs and ICE vehicles across all tested delay durations. In contrast, no collisions were observed in the other scenarios.
This scenario's heightened susceptibility stems from the adjacent positioning of attacked vehicles (vehicles 5 and 6), which amplifies the destabilizing effects of discretized information delays. Under the 6-second delay condition, collisions manifested at 85.73 seconds for EVs and 80.23 seconds for ICE vehicles, indicating that ICE vehicles exhibit greater vulnerability to DPDA-induced instabilities. Moreover, extending the delay to 8 and 9 seconds precipitated earlier collision events, demonstrating a negative correlation between delay duration and system stability.

\begin{figure}[htbp]
\centering
\subfloat[EV spacing.\label{fig:ev-spacing-dpda}]{
    \includegraphics[width=0.45\linewidth]{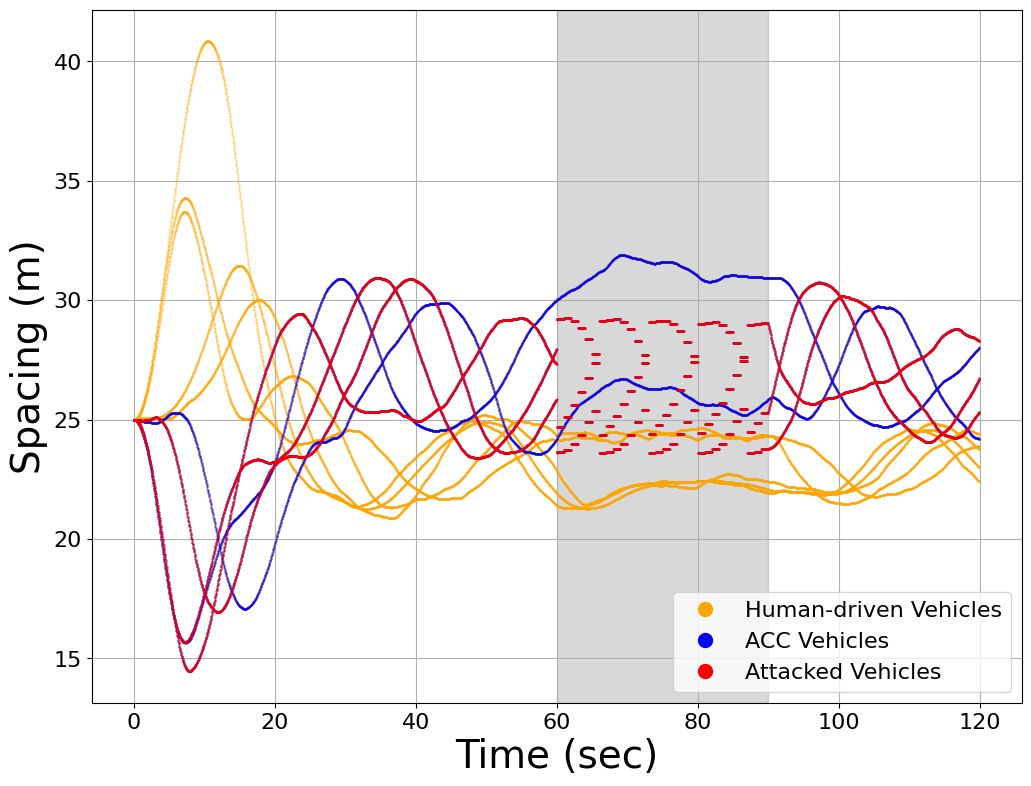}
}
\hfill
\subfloat[EV speed.\label{fig:ev-speed-dpda}]{
    \includegraphics[width=0.45\linewidth]{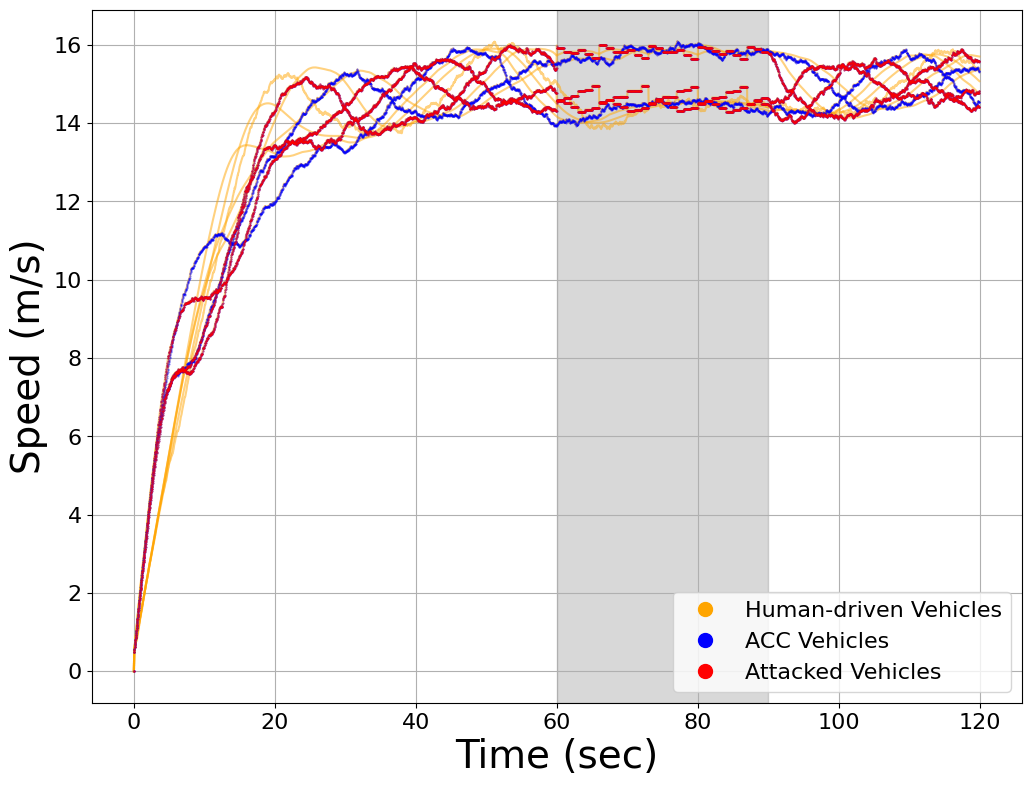}
}

\vspace{0em}  

\subfloat[ICE vehicle spacing.\label{fig:ice-spacing-dpda}]{
    \includegraphics[width=0.45\linewidth]{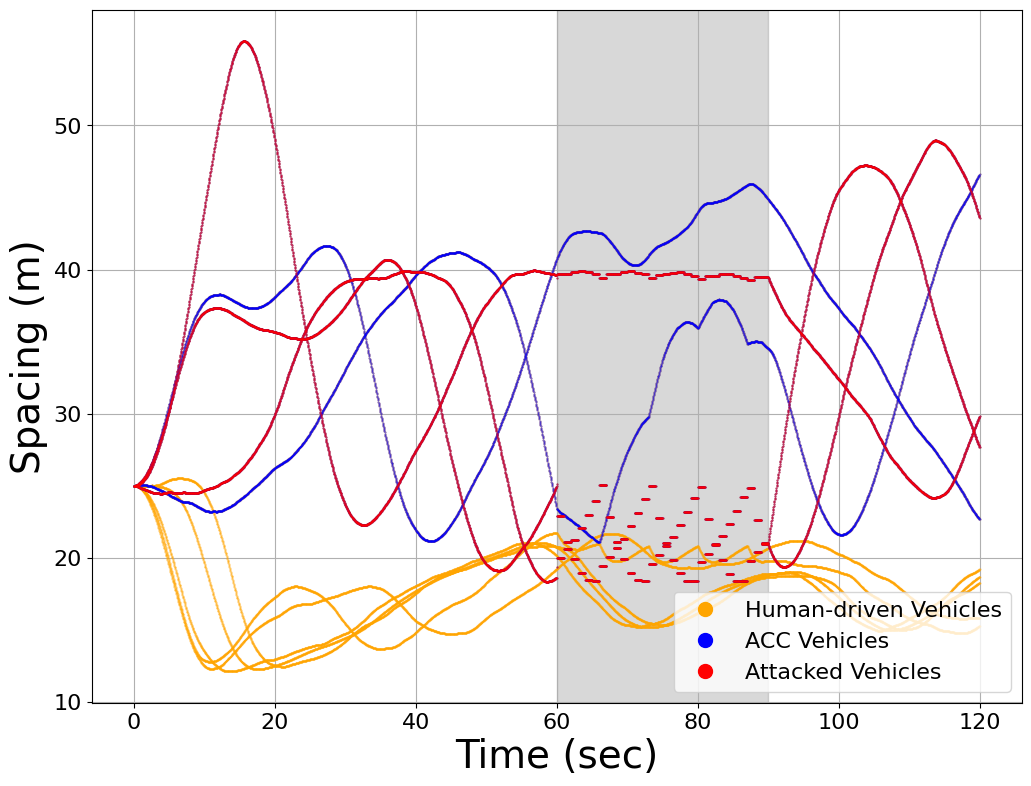}
}
\hfill
\subfloat[ICE vehicle speed.\label{fig:ice-speed-dpda}]{
    \includegraphics[width=0.45\linewidth]{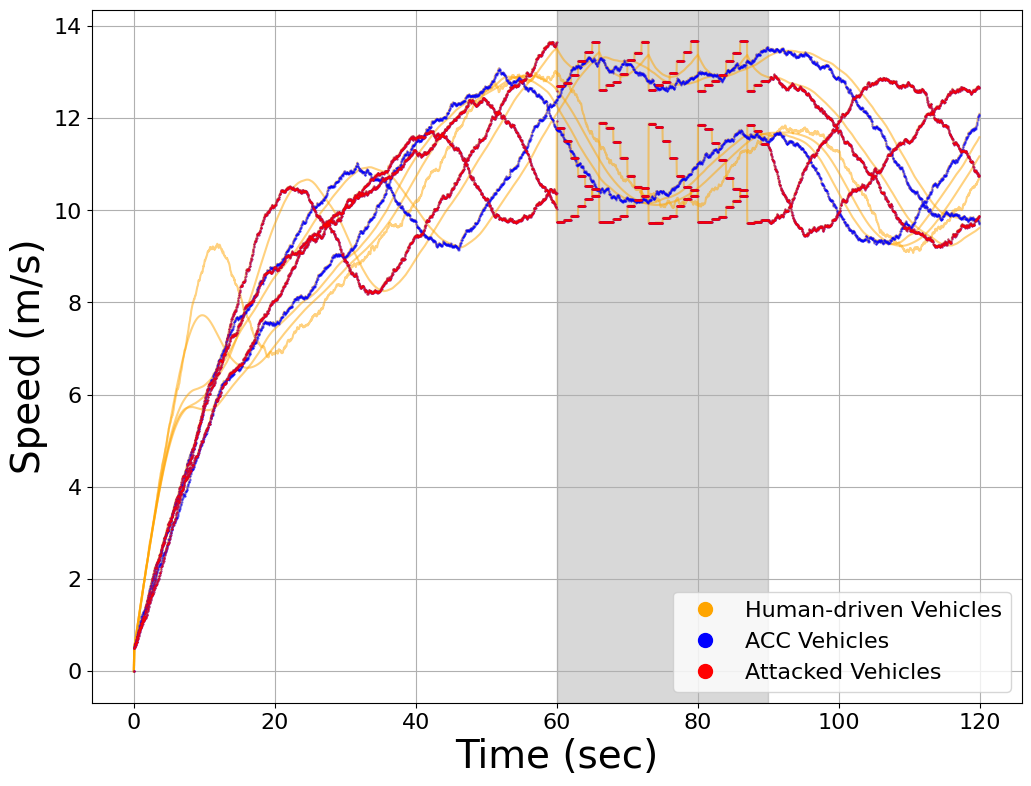}
}

\caption{Comparative analysis of spacing and velocity dynamics between EVs and ICE vehicles under DPDA in Scenario IV (6-second delay).}
\label{fig:dpda-dynamics}
\end{figure}

Fig. \ref{fig:dpda-dynamics} illustrates the collision dynamics for Scenario IV under the 6-second delay condition, representing the least severe case among the observed collision scenarios. The comparative analysis reveals distinct failure modes: EVs maintain higher operational velocities until the onset of the collision (Fig. \ref{fig:ev-speed-dpda} vs. \ref{fig:ice-speed-dpda}), while ICE vehicles exhibit more pronounced spacing fluctuations prior to the collision event (Fig. \ref{fig:ice-spacing-dpda} vs. \ref{fig:ev-spacing-dpda}). 
Due to the occurrence of collisions during the attack phase, performance metrics such as $\overline{SSD}$ and $\overline{VSD}$ in Table \ref{under six cyberattacks} are rendered invalid for both the during-attack and post-attack phases. As a result, these values are excluded from the corresponding records.

\begin{table}[h]
\caption{\\Collision status across simulation scenarios under DPDA with varying delay time. Bold entries indicate collision occurrence.}
\label{Combinations of Delay Times and Scenarios}
\centering
\begin{tabular}{ccccc}
\hline
\multirow{2}{*}{\textbf{Delay Time (s)}} & \multicolumn{4}{c}{\textbf{Simulation Scenarios}} \\ \cline{2-5}
& I & II & III & IV   \\ \hline
6                           & No  & No  & No  & \textbf{Yes} \\
8                           & No  & No  & No  & \textbf{Yes} \\
9                           & No  & No  & No  & \textbf{Yes} \\ \hline
\end{tabular}
\end{table}

\subsection{Phantom Attack Analysis}
To maintain experimental tractability, the PA implementation assumes that each compromised vehicle receives information intended for its immediate predecessor in the platoon. This configuration represents a realistic attack scenario where malicious actors redirect V2V communication streams between consecutive vehicles.

The attack configuration can be represented through a modified adjacency matrix $\widetilde{\bm{\mathrm{M}}}$, where non-zero entries indicate corrupted information flows. For instance, the entry $\widetilde{\bm{\mathrm{M}}}(3,1) = 1$ indicates that vehicle 3 receives information originally intended for vehicle 1, thereby creating a mismatch between perceived and actual traffic conditions:

\begin{equation*}
  \widetilde{\bm{\mathrm{M}}}  =
\begin{bmatrix}
    0 & 0 & \cdots & 0 &  \\
    0& 0 & \cdots & 0 &\\
     \textbf{1}& 0 & \cdots & 0\\
   \vdots & \vdots     & \cdots     &  \vdots   &       \\
    0 & 0 & \cdots & 0 &  \\
\end{bmatrix}_{10\times 10}
\end{equation*}

Simulation results across all four scenarios indicate that, despite introducing substantial disturbances, PA does not trigger collision events for either EVs or ICE vehicles. Therefore, the analysis presented herein focuses exclusively on the worst-case configuration, namely Scenario IV with adjacent attacked vehicles. The observed resilience is primarily attributed to the inherent stability margins embedded within ACC systems and the bounded nature of the falsified information, which remains within physically plausible limits, and the relatively steady vehicle speeds in the experimental environment.

\begin{figure}[htbp]
\centering
\subfloat[EV spacing.\label{fig:ev-spacing-phantom}]{
    \includegraphics[width=0.45\linewidth]{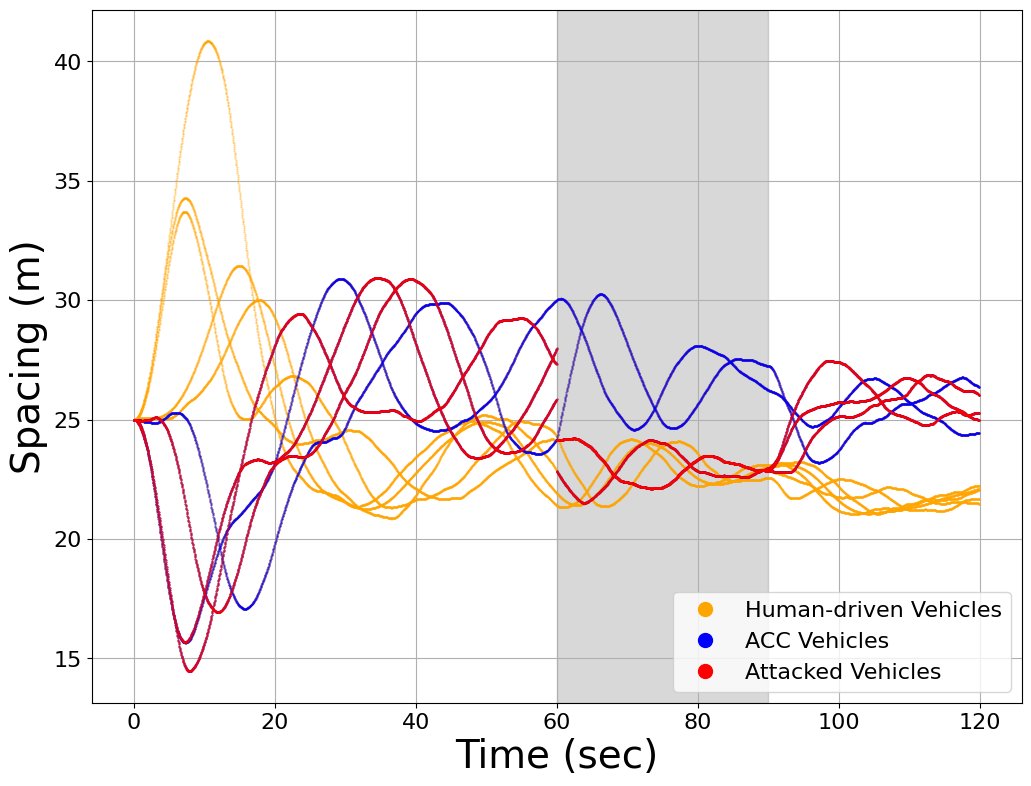}
}
\hfill
\subfloat[EV speed.\label{fig:ev-speed-phantom}]{
    \includegraphics[width=0.45\linewidth]{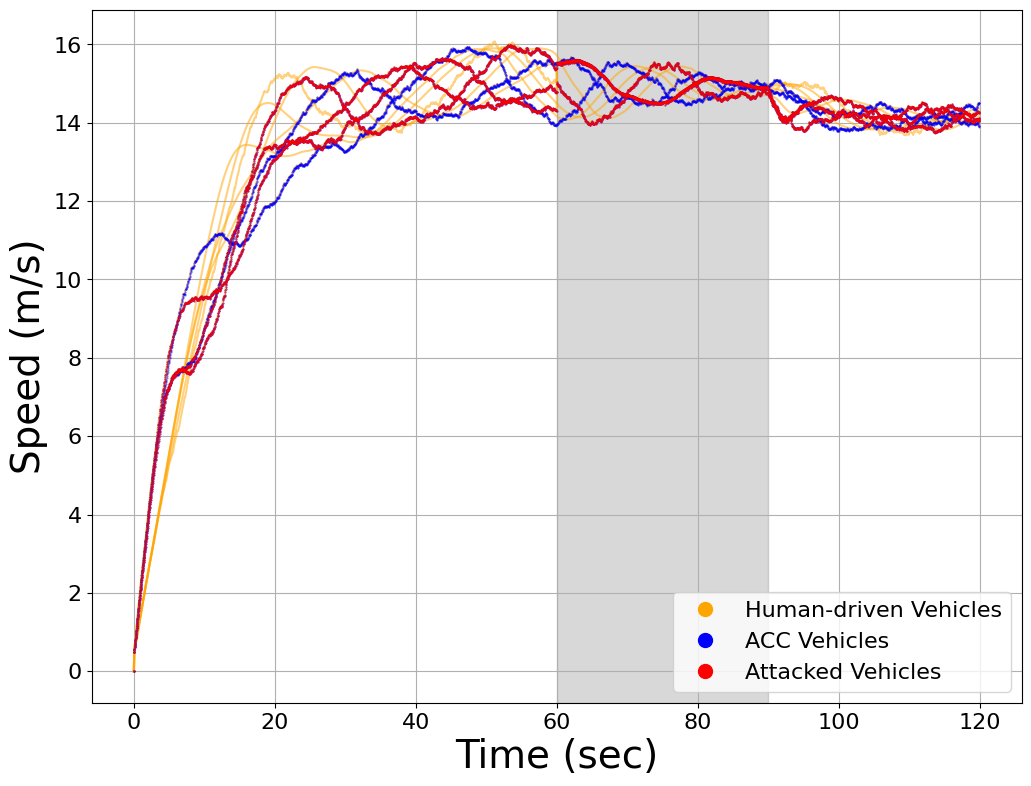}
}

\vspace{0em}  

\subfloat[ICE vehicle spacing.\label{fig:ice-spacing-phantom}]{
    \includegraphics[width=0.45\linewidth]{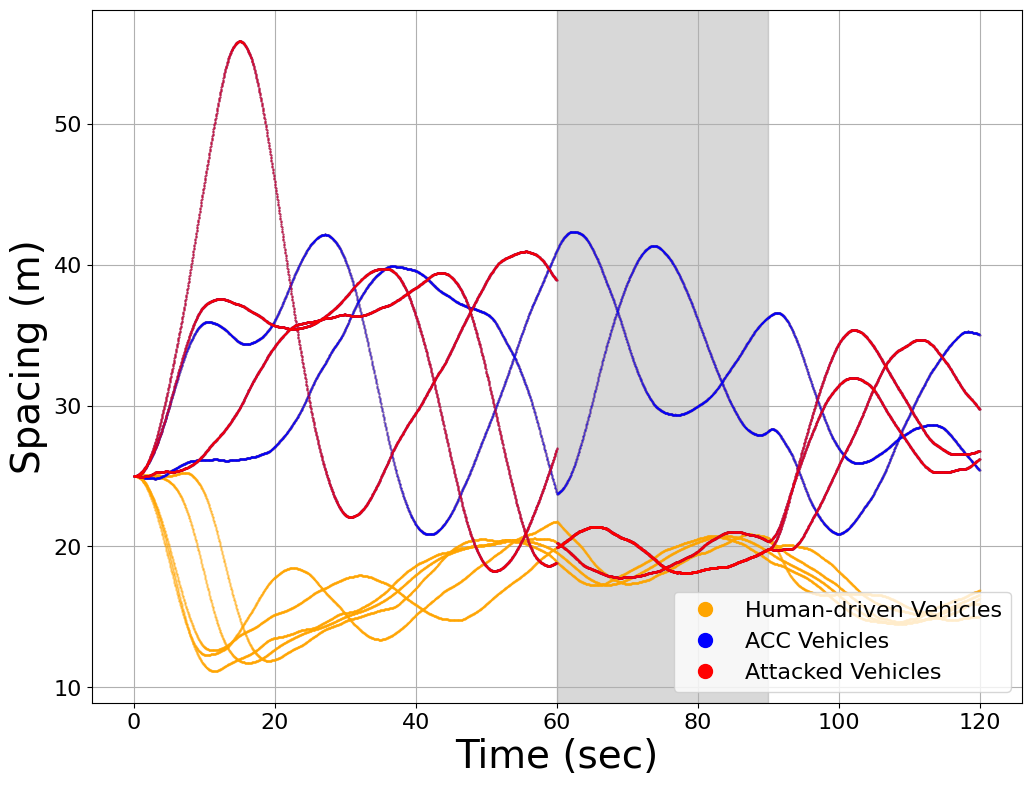}
}
\hfill
\subfloat[ICE vehicle speed.\label{fig:ice-speed-phantom}]{
    \includegraphics[width=0.45\linewidth]{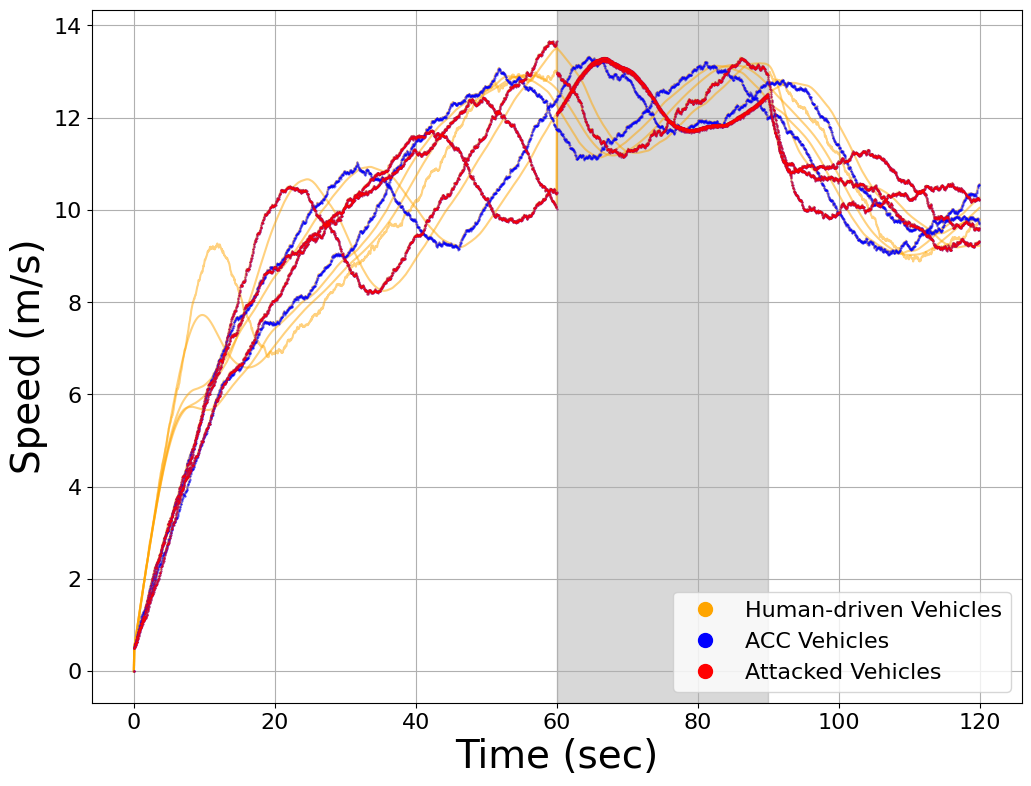}
}

\caption{Comparative analysis of spacing and velocity dynamics between EVs and ICE vehicles under PA in Scenario IV.}
\label{fig:phantom-dynamics}
\end{figure}
Despite the absence of collisions, distinct behavioral differences emerge between EVs and ICE vehicles in Fig. \ref{fig:phantom-dynamics}. EVs demonstrate superior stability characteristics, maintaining more consistent spacing patterns (Fig. \ref{fig:ev-spacing-phantom} vs. \ref{fig:ice-spacing-phantom}) and exhibiting reduced velocity fluctuations (Fig. \ref{fig:ev-speed-phantom} vs. \ref{fig:ice-speed-phantom}) compared to ICE vehicles.
Quantitative analysis in Table \ref{under six cyberattacks} corroborates these observations. EVs consistently exhibit lower $\overline{SSD}$ and $\overline{VSD}$ values across all simulation phases, indicating superior disturbance rejection capabilities. Furthermore, EVs maintain reduced $THW$ and elevated $V_{avg}$ values compared to ICE vehicles, reflecting their ability to sustain tighter and faster platoon formations without compromising stability. These findings suggest that the superior acceleration and responsive control capabilities of EVs confer greater resilience against information-manipulation attacks (e.g., PA).

\subsection{Fixed-Speed Attack Analysis}

To evaluate the impact of FA on platoon dynamics, simulations were configured such that compromised vehicles maintain their instantaneous velocity at the attack initiation time (t = 60 seconds) throughout the entire attack phase. This constraint effectively disables the ACC's adaptive velocity control, forcing vehicles to operate at constant speed regardless of changing traffic conditions.

Experimental results reveal critical vulnerabilities in Scenario IV, where adjacent positioning of attacked vehicles (vehicles 5 and 6) creates a cascading failure mode. Notably, the absence of collisions in the other three scenarios led to the selection of Scenario IV for detailed analysis.
In Scenario IV, both EV and ICE platoons experienced collision events, albeit with distinct temporal characteristics. ICE vehicles demonstrated higher vulnerability, with collision onset at 75.8 seconds, whereas EVs exhibited enhanced resilience with collision occurrence delayed until 80.23 seconds. This 4.43-second differential highlights the superior disturbance rejection capabilities of EVs.

\begin{figure}[htbp]
\centering
\subfloat[EV spacing.]{
    \includegraphics[width=0.45\linewidth]{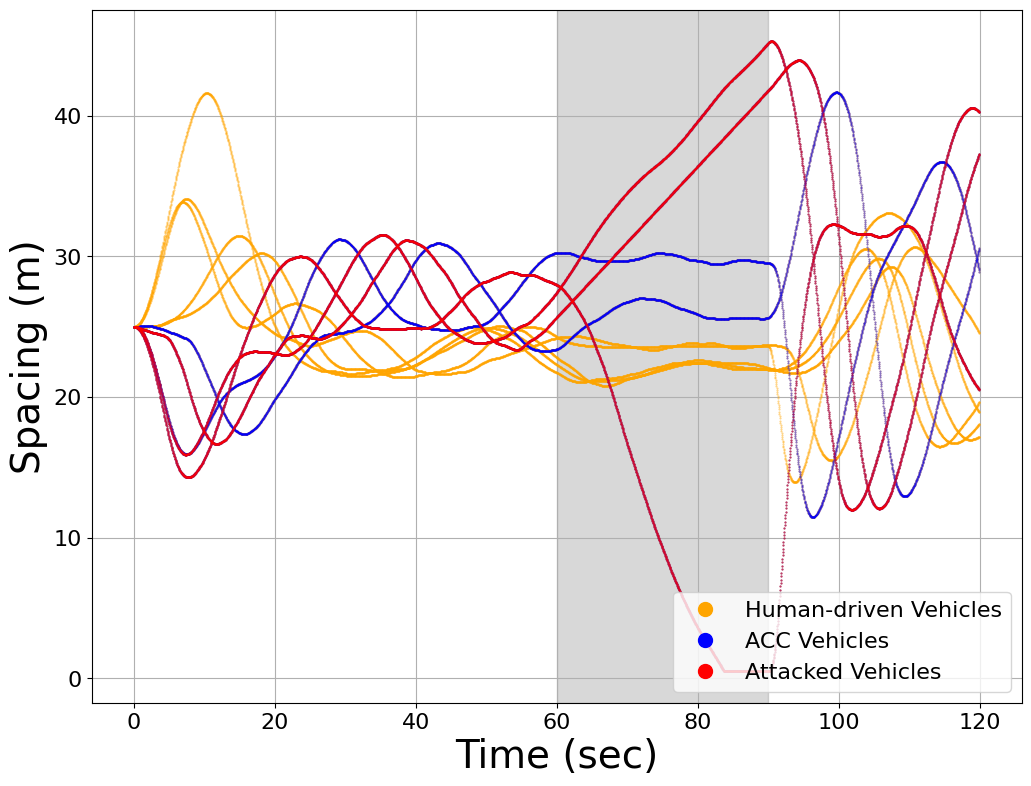}
    \label{fig:ev-spacing-fixed}
}
\hfill
\subfloat[EV speed.]{
    \includegraphics[width=0.45\linewidth]{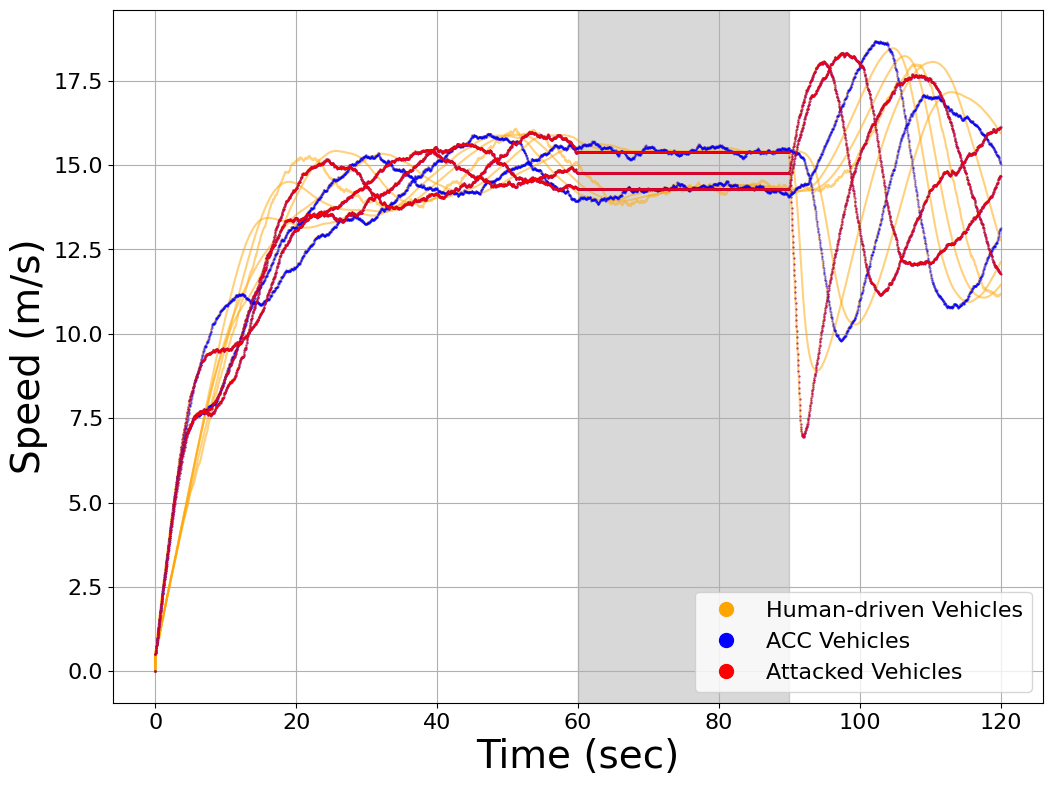}
    \label{fig:ev-speed-fixed}
}

\vspace{0em}

\subfloat[ICE vehicle spacing.]{
    \includegraphics[width=0.45\linewidth]{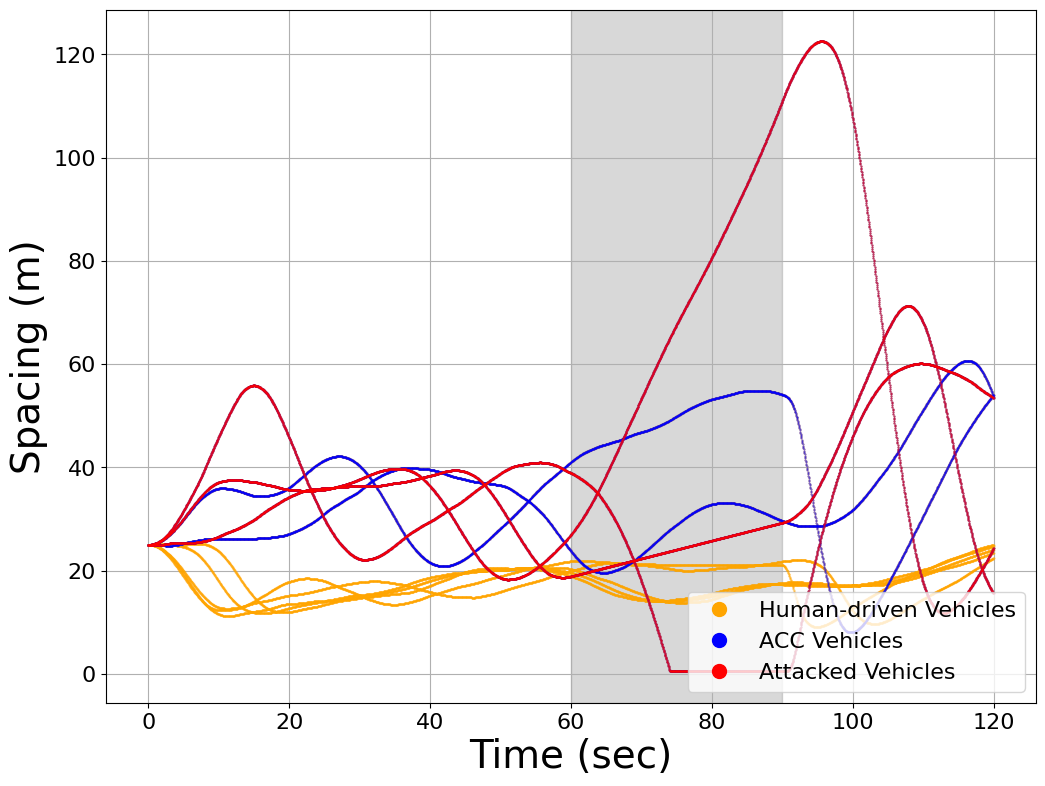}
    \label{fig:ice-spacing-fixed}
}
\hfill
\subfloat[ICE vehicle speed.]{
    \includegraphics[width=0.45\linewidth]{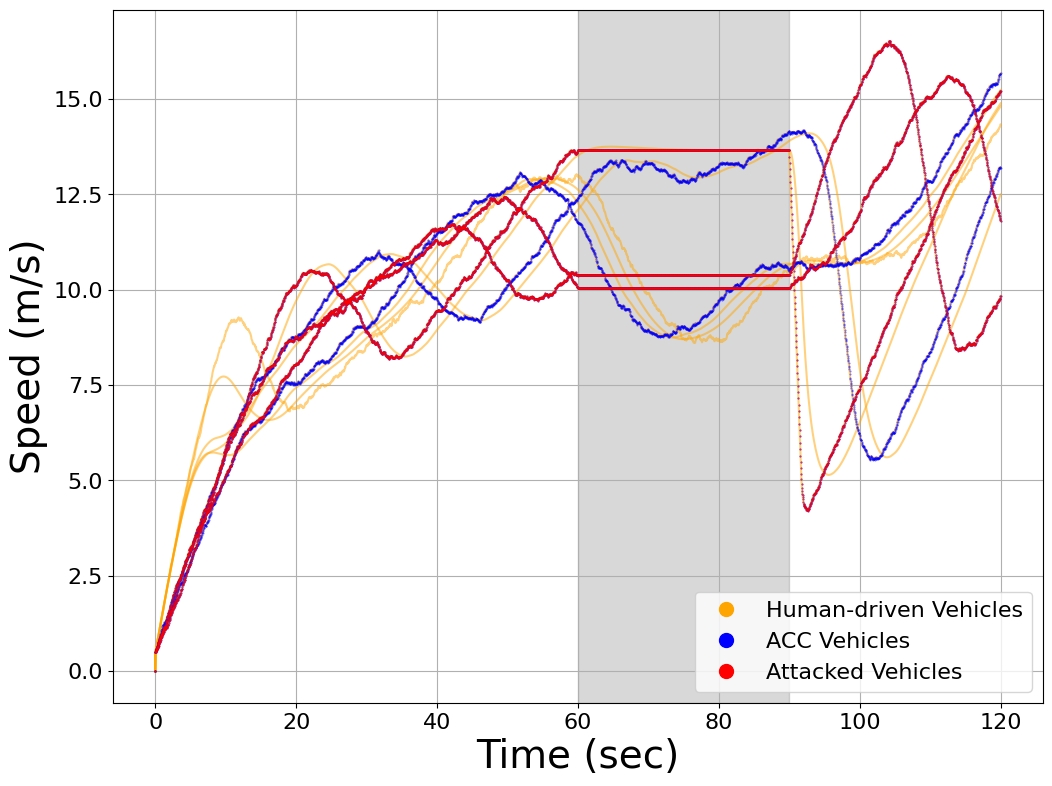}
    \label{fig:ice-speed-fixed}
}

\caption{Comparative analysis of spacing and velocity dynamics between EVs and ICE vehicles under FA in Scenario IV.}
\label{Figurefixed attack}
\end{figure}

Fig. \ref{Figurefixed attack} illustrates the collision dynamics for Scenario IV. Analysis of the velocity profiles reveals that EVs maintained higher operational speeds and more aggressive following behavior until collision onset (Fig. \ref{fig:ev-speed-fixed} vs. \ref{fig:ice-speed-fixed}), effectively utilizing their superior acceleration capabilities to maintain platoon cohesion. Conversely, ICE vehicles exhibited conservative spacing strategies with larger inter-vehicle gaps (Fig. \ref{fig:ice-spacing-fixed} vs. \ref{fig:ev-spacing-fixed}), yet paradoxically experienced earlier collision events due to their limited acceleration response when traffic conditions demanded rapid adaptation.
Owing to the collision occurring during the attack phase, metrics such as $\overline{SSD}$ and $\overline{VSD}$ reported in Table \ref{under six cyberattacks} lose their interpretive significance during both the during-attack and post-attack phases, and are therefore excluded from the results.

The quantitative metrics presented in Table \ref{under six cyberattacks} corroborate these observations. Pre-collision analysis indicates that EVs maintained lower $\overline{VSD}$ and $\overline{SSD}$ values, demonstrating superior stability under attack conditions. Furthermore, the reduced $THW$ values for EVs reflect their ability to sustain efficient platoon formations even when ACC functionality is compromised. These findings underscore a critical insight: while EVs' aggressive following behavior might appear riskier, their superior dynamic response capabilities ultimately provide enhanced resilience against FA compared to the more conservative but dynamically limited ICE platforms.

\subsection{Blinding Attack Analysis}
Safe following distances in vehicular operations are proportionally scaled with vehicle speed to ensure sufficient reaction time and braking distance, thereby maintaining operational safety under varying traffic conditions.
This study concentrates on Scenario I with a simplified attack configuration of a single compromised vehicle. This setup enables the isolation and examination of the fundamental attack dynamics without interference from multiple attack vectors. Given that collisions are observed under this minimal-impact Scenario I in Table \ref{tab:cyberattack_performance}, incidents would also manifest in other complex simulation scenarios. Hence, the BA analysis is restricted to Scenario I for clarity and conciseness.

Kim et al. \cite{kim2021development} suggest maintaining a minimum inter-vehicle distance of 50 meters to ensure safe driving at a speed of 60 km/h. Considering $V_{avg}$ during the attack phase is 15.74 m/s (i.e., 56.66 km/h) in Table \ref{no attack scenarios}, this study adopts the recommended 50-meter threshold to define the maximum perceived spacing ($\varphi = 50$) for both EVs and ICE vehicles in Eq. \ref{eq:BA}, thereby aligning with established traffic safety guidelines.
Noteworthy, this study adopts two preceding vehicles for BA such that the cumulative distance from vehicle $1$ to vehicle $8$ is given by: 
\begin{equation}
   \sum_{z=0}^{2}s_{(1-z)\bmod 10} = s_1+ s_0 + s_9 = 75 
\end{equation}
Since the cumulative distance exceeds the threshold $\varphi = 50$, the perceived spacing $\tilde{s} = \varphi $ is used in Eq. \ref{eq:BA} of BA.

Simulation results demonstrate that BA triggers collision events across all simulation scenarios for both EVs and ICE vehicles, underscoring the severe threat posed by perception-manipulation attacks. 
In particular, collisions occurred at different time steps: $63.03$ seconds for EVs and $88.83$ seconds for ICE vehicles in Scenario I. This 25.8-second discrepancy highlights a counterintuitive vulnerability pattern in which EVs, despite their inherently superior dynamic capabilities, exhibit increased susceptibility to BA. This reversal of the typical EV advantage may require further in-depth investigation.

\begin{figure}[htbp]
\centering
\subfloat[EV spacing.]{
    \includegraphics[width=0.45\linewidth]{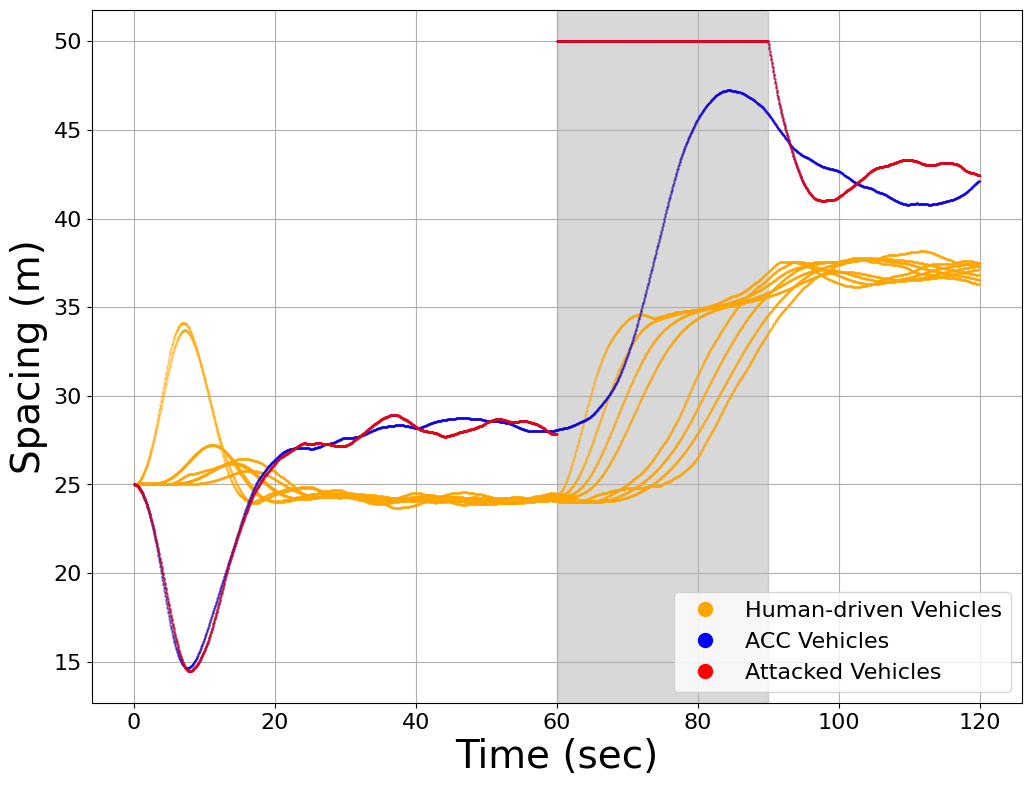}
    \label{ev-spacing-blinding}
}
\hfill
\subfloat[EV speed.]{
    \includegraphics[width=0.45\linewidth]{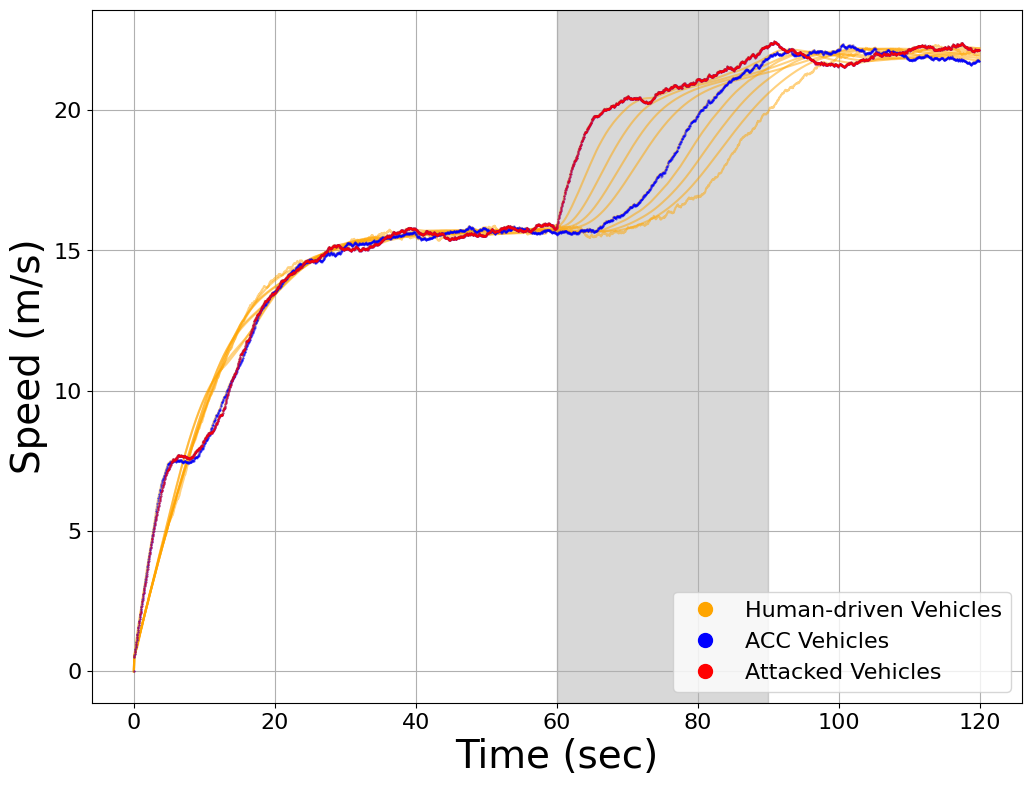}
    \label{ev-speed-blinding}
}

\vspace{0em}

\subfloat[ICE vehicle spacing.]{
    \includegraphics[width=0.45\linewidth]{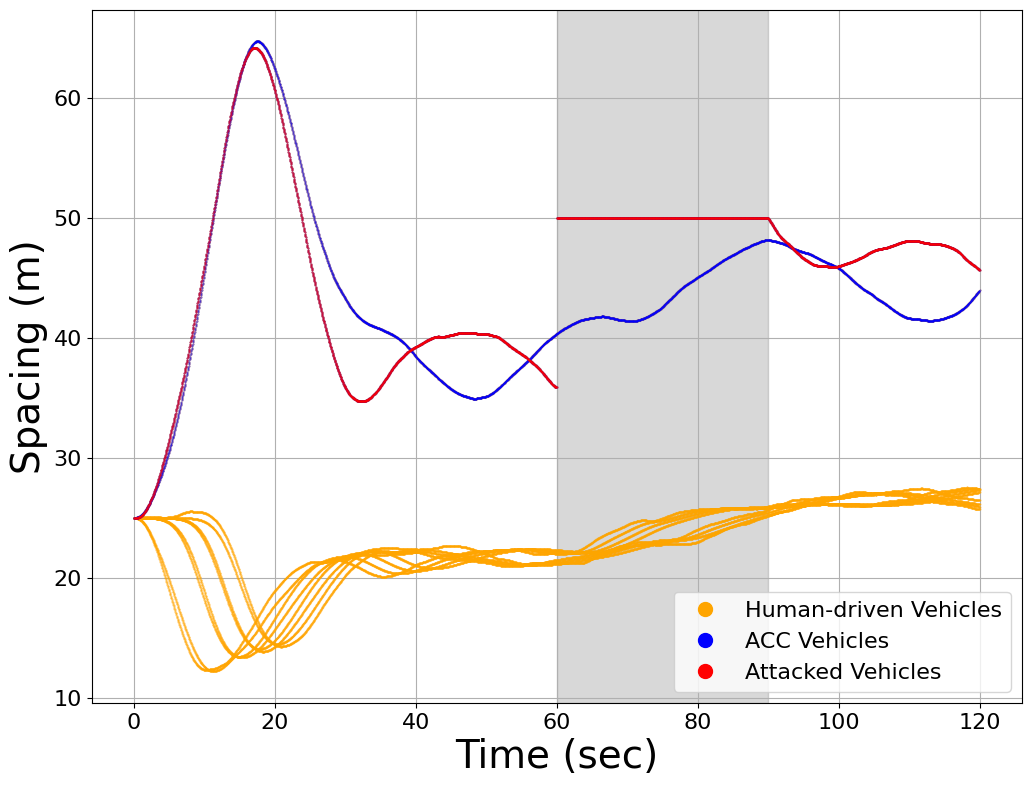}
    \label{ice-spacing-blinding}
}
\hfill
\subfloat[ICE vehicle speed.]{
    \includegraphics[width=0.45\linewidth]{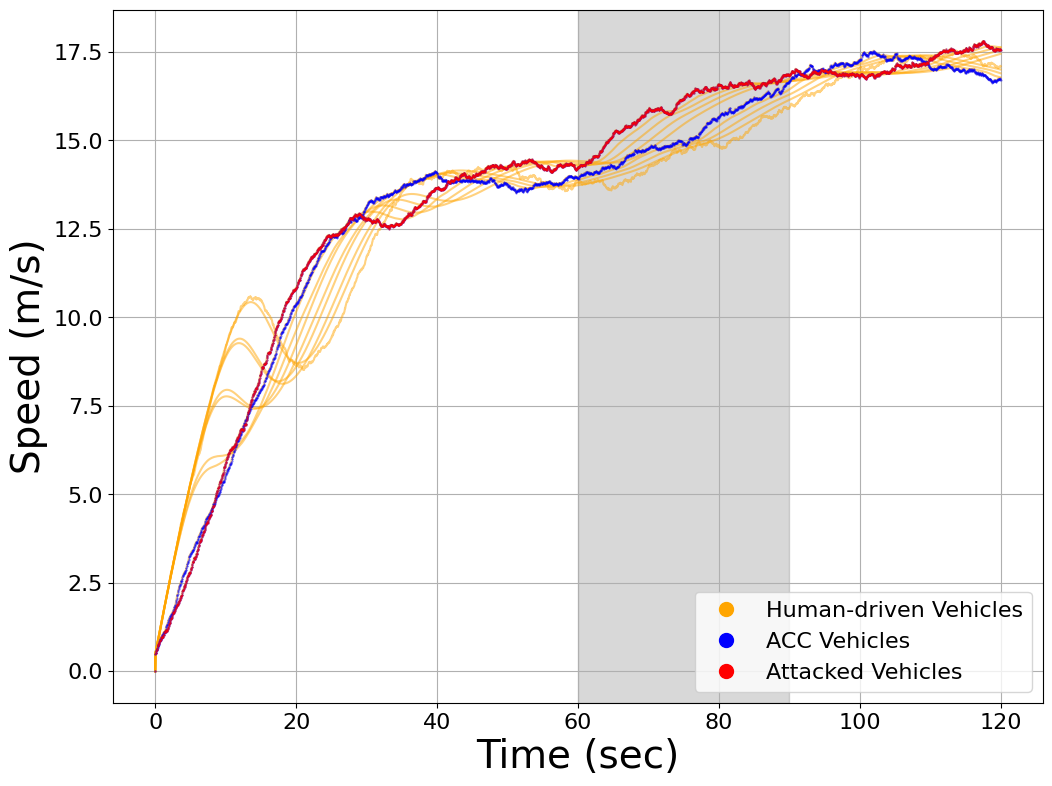}
    \label{ice-speed-blinding}
}

\caption{Comparative analysis of spacing and velocity dynamics between EVs and ICE vehicles under BA in Scenario I.}
\label{Figureblinding attack}
\end{figure}

Fig. \ref{Figureblinding attack} illustrates the collision dynamics, highlighting distinct failure modes between the two vehicle types. EVs consistently maintain higher velocities during the pre-collision phase (Fig. \ref{ev-speed-blinding} vs. \ref{ice-speed-blinding}), whereas ICE vehicles exhibit larger inter-vehicle spacing (Fig. \ref{ice-spacing-blinding} vs. \ref{ev-spacing-blinding}), indicating divergent behavioral responses to attack-induced disruptions. This divergence becomes particularly critical under BA, where the compromised perception system fails to recognize intermediate vehicles.
It should be noted that performance results for BA during the attack and post-attack phases are missing in Table \ref{under six cyberattacks} due to the occurrence of collision events.

\begin{table*}[h] 
\caption{\\ Comparative performance of EVs and ICE vehicles under various proposed cyberattacks within a specific simulation scenario. The symbol ‘---’ denotes missing data resulting from collision events. Units: $V_{avg}$ (m/s), $\overline{VSD}$ (m/s), $\overline{SSD}$ (m), $THW$ (s).}
\label{tab:cyberattack_performance}
\centering
\resizebox{\textwidth}{!}{
\begin{tabular}{@{}
>{\centering\arraybackslash}m{1.5cm}
>{\centering\arraybackslash}m{1.5cm}
>{\centering\arraybackslash}m{1.5cm}
*{6}{S[table-format=2.2, table-column-width=1.5cm]} @{}} 
\toprule
\multirow{2}{*}{\textbf{Attack}}&\multirow{2}{*}{\textbf{Scenario}}&\multirow{2}{*}{\textbf{Metric}} & \multicolumn{3}{c}{\textbf{EV}} & \multicolumn{3}{c}{\textbf{ICE vehicle}} \\
\cmidrule(lr){4-6} \cmidrule(lr){7-9}
& & & {\textbf{Pre}} & {\textbf{During}} & {\textbf{Post}} & {\textbf{Pre}} & {\textbf{During}} & {\textbf{Post}} \\
\midrule
\multirow{4}{*}{ DPDA} &\multirow{4}{*}{ IV}  & \textbf{$V_{avg}$} & 14.82 & {---} & {---} & 10.96 & {---} & {---} \\
               &         &  $\overline{VSD}$            &  0.47 & {---} & {---} &  1.10 & {---} & {---} \\
               &         & $\overline{SSD}$          &  1.76 & {---} & {---} &  3.21 & {---} & {---} \\
                &        &$ THW$              &  1.68 & {---} & {---} &  2.30 & {---} & {---} \\
\midrule
\multirow{4}{*}{PA} &\multirow{4}{*}{ IV}    & \textbf{$V_{\text{avg}}$} & 14.88 & 14.87 & 14.16 & 11.08 & 12.27 & 10.38 \\
                &        & $\overline{VSD}$             &  0.47 &  0.21 &  0.32 &  1.10 &  0.52 &  1.11 \\
                 &       & $\overline{SSD}$           &  1.15 &  0.90 &  0.84 &  3.69 &  1.64 &  2.69 \\
                 &       & $ THW$             &  1.68 &  1.59 &  1.66 &  2.30 &  1.79 &  2.16 \\
\midrule
\multirow{4}{*}{FA}  &\multirow{4}{*}{ IV}   & \textbf{$V_{\text{avg}}$} & 14.88 & {---} & {---} & 11.08 & {---} & {---} \\
              &          & $\overline{VSD}$        &  0.47 & {---} & {---} &  1.10 & {---} & {---} \\
               &         & $\overline{SSD}$           &  1.15 & {---} & {---} &  3.69 & {---} & {---} \\
               &         & $ THW$              &  1.68 & {---} & {---} &  2.30 & {---} & {---} \\
\midrule
\multirow{4}{*}{BA}  &\multirow{4}{*}{ I}   & \textbf{$V_{\text{avg}}$} & 15.59 & {---} & {---} & 13.78 & {---} & {---} \\
               &         &$\overline{VSD}$               &  0.20 & {---} & {---} &  0.38 & {---} & {---} \\
               &        &$\overline{SSD}$              &  0.34 & {---} & {---} &  0.95 & {---} & {---} \\
               &         & $ THW$               &  1.60 & {---} & {---} &  1.83 & {---} & {---} \\
\midrule
\multirow{4}{*}{AVA} &\multirow{4}{*}{ IV}   & \textbf{$V_{\text{avg}}$} & 14.88 & 15.49 & 14.96 & 11.08 & 11.84 & 11.83 \\
              &          & $\overline{VSD}$            &  0.47 &  0.61 &  0.52 &  1.10 &  1.24 &  1.91 \\
               &         & $\overline{SSD}$           &  1.15 &  2.30 &  1.80 &  3.69 &  5.18 &  6.21 \\
              &          & $ THW$         &  1.68 &  1.61 &  1.67 &  2.30 &  2.08 &  2.10 \\
\midrule
\multirow{4}{*}{MA} &\multirow{4}{*}{ IV}    & \textbf{$V_{\text{avg}}$} & 14.88 & 14.98 & 14.58 & 11.08 & 11.14 & 10.07 \\
               &         & $\overline{VSD}$             &  0.47 &  0.38 &  0.48 &  1.10 &  1.11 &  2.05 \\
              &          & $\overline{SSD}$             &  1.15 &  1.31 &  1.40 &  3.69 &  3.16 &  6.45 \\
              &          &$ THW$               &  1.68 &  1.60 &  1.66 &  2.30 &  1.91 &  2.16 \\
\bottomrule
 \label{under six cyberattacks}
\end{tabular}
}
\end{table*}

\subsection{Angular Velocity Attack Analysis}
To ensure experimental tractability, the function $G(\phi) = 1 + k \cdot \sin(\phi)$ is defined in Eq. \ref{AVA-eq}, where $k$ denotes the interference coefficient that modulates the intensity of the attack. A value of $k = 0.002$ is selected to induce gradual variations in impact, thereby providing a more realistic approximation of real-world attack dynamics.

Simulation results indicate that AVA does not trigger collision events in any of the simulation scenarios for either vehicle type.
Even under the most critical configuration, namely Scenario IV with adjacent attacked vehicles indexed by 5 and 6, both EVs and ICE vehicles maintain collision-free operation throughout the during-attack phase. Consequently, results for less severe configurations in other simulation scenarios are not presented for brevity.
This observed resilience can be attributed to the relatively minor velocity perturbations induced by the selected interference coefficient, which remain within the adaptive capacity of the ACC systems.

\begin{figure}[htbp]
\centering
\subfloat[EV spacing.]{
    \includegraphics[width=0.45\linewidth]{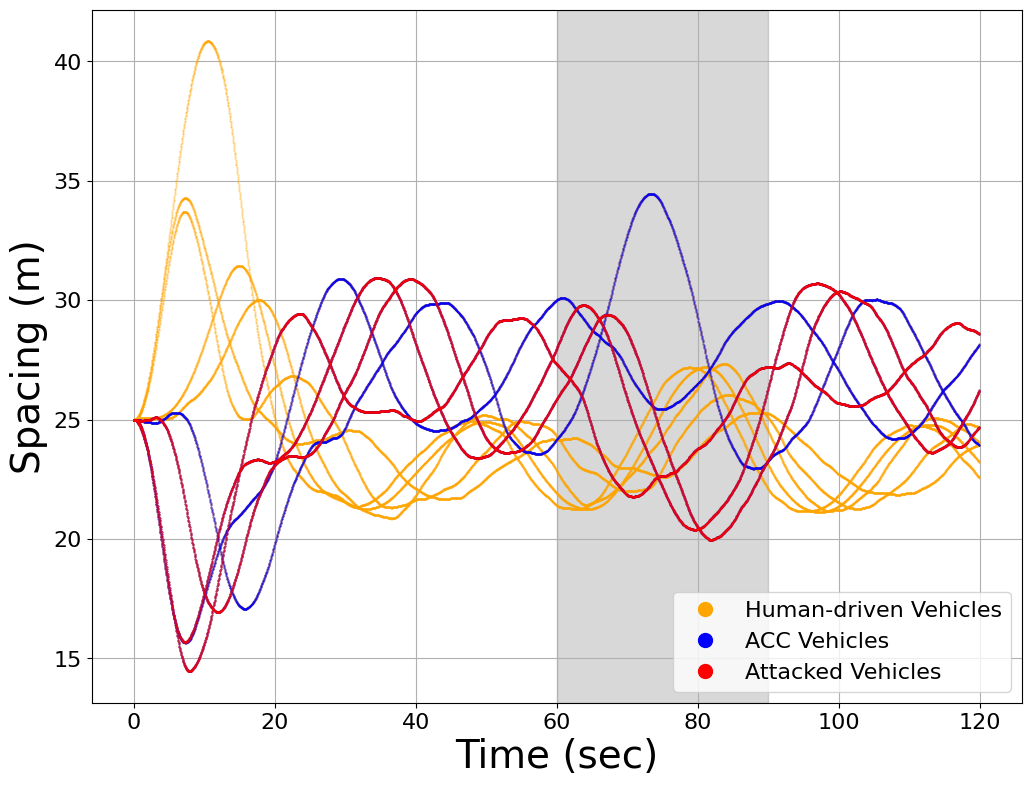}
    \label{ev-spacing-angular}
}
\hfill
\subfloat[EV speed.]{
    \includegraphics[width=0.45\linewidth]{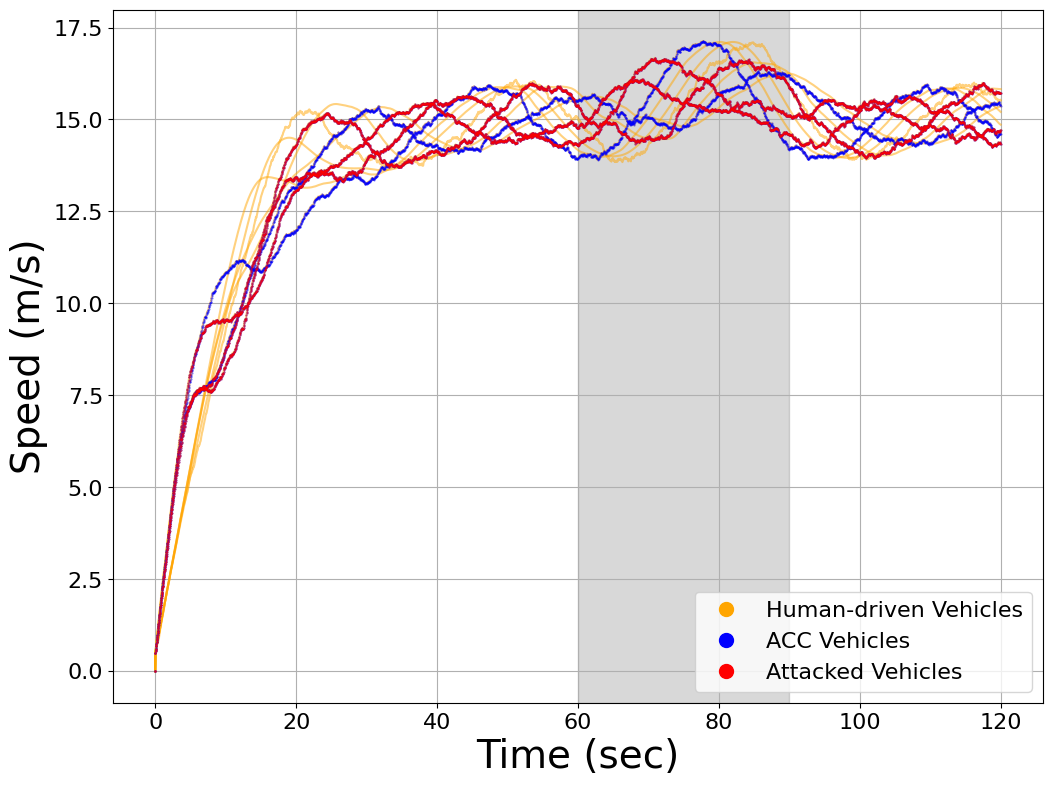}
    \label{ev-speed-angular}
}

\vspace{0em}

\subfloat[ICE vehicle spacing.]{
    \includegraphics[width=0.45\linewidth]{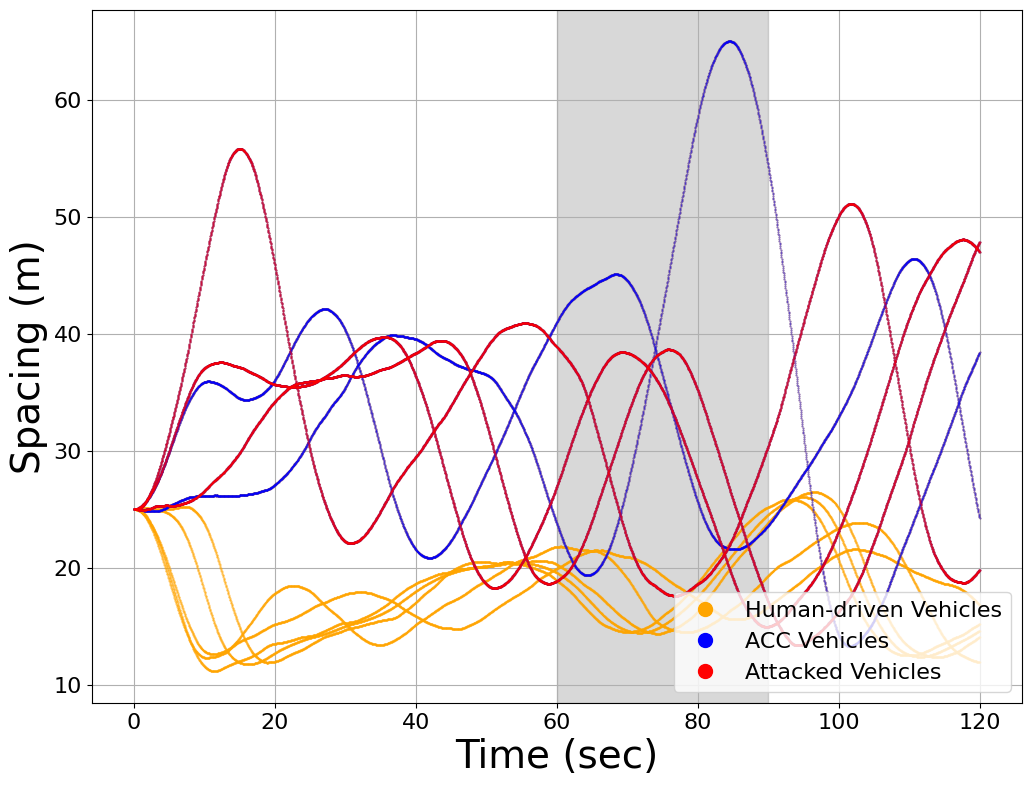}
    \label{ice-spacing-angular}
}
\hfill
\subfloat[ICE vehicle speed.]{
    \includegraphics[width=0.45\linewidth]{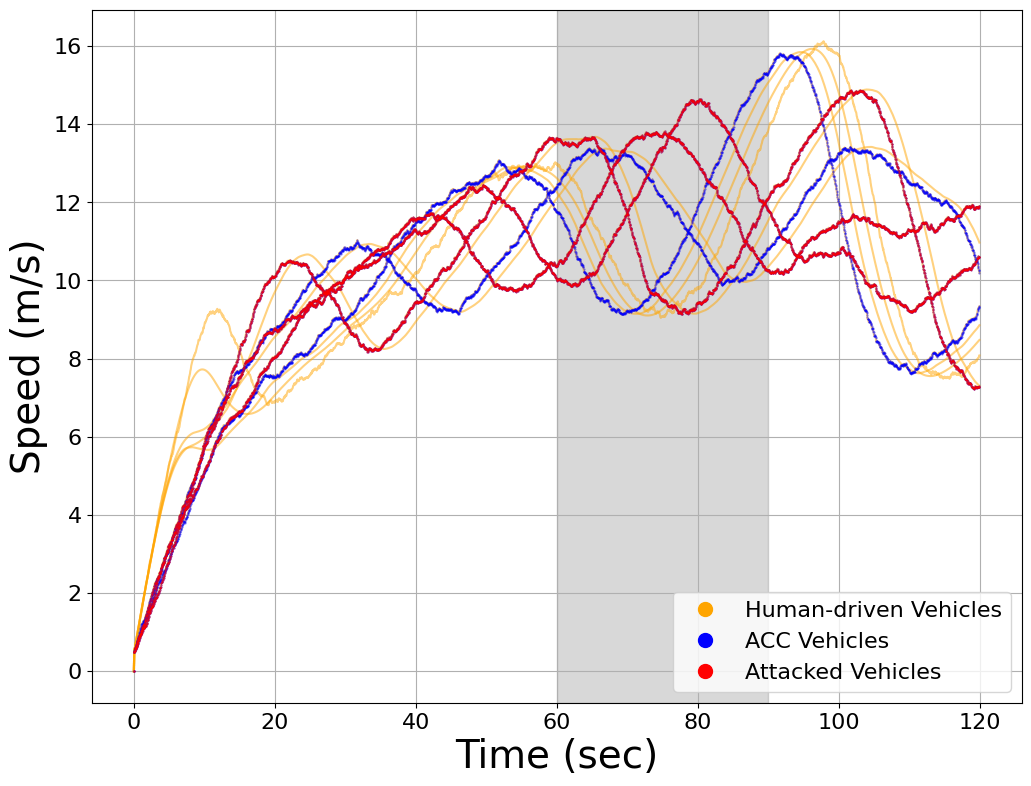}
    \label{ice-speed-angular}
}

\caption{Comparative analysis of spacing and velocity dynamics between EVs and ICE vehicles under AVA in Scenario IV.}
\label{Figureangular velocity attack}
\end{figure}

Fig. \ref{Figureangular velocity attack} presents the dynamic response in Scenario IV, revealing distinct behavioral patterns between two vehicle types. EVs consistently maintain higher operational velocities (Fig. \ref{ev-speed-angular} vs. \ref{ice-speed-angular}) and exhibit reduced inter-vehicle spacing (Fig. \ref{ev-spacing-angular} vs. \ref{ice-spacing-angular}), indicating superior platoon stability. In contrast, ICE vehicles exhibit more pronounced spacing fluctuations and decreased average speeds, indicating degraded platoon cohesion under angular velocity perturbations.

Quantitative analysis in Table \ref{under six cyberattacks} corroborates these observations. EVs exhibit lower $THW$ and enhanced $V_{avg}$ compared to ICE vehicles, indicating their ability to maintain faster and more compact platoon formations.
Despite the absence of collision events, ICE vehicles demonstrate significantly elevated $\overline{VSD}$ and $\overline{SSD}$ values, particularly during post-attack phases, indicating compromised recovery capabilities. 
This disparity highlights the inherent advantage of EVs in handling angular velocity disturbances. Their superior torque response and control precision facilitate more effective compensation for velocity perturbations, thereby maintaining platoon stability where ICE vehicles struggle to recover baseline performance.

\subsection{Mixed Attack Analysis}

The MA experimental framework integrates key characteristics from both PA and DPDA. In particular, compromised vehicles receive information intended for their predecessors (PA component), while simultaneously subjected to discretized temporal delays (DPDA component). Simulations were performed using the same delay intervals employed in the DPDA experiments, as documented in Table \ref{Combinations of Delay Times and Scenarios}. This study focuses on the most critical configuration, namely Scenario IV with adjacent compromised vehicles (vehicles 5 and 6) and a delay parameter of $9$ seconds. Accordingly, results from less severe configurations in alternative scenarios are omitted for brevity.

\begin{figure}[htbp]
\centering
\subfloat[EV spacing.]{
    \includegraphics[width=0.45\linewidth]{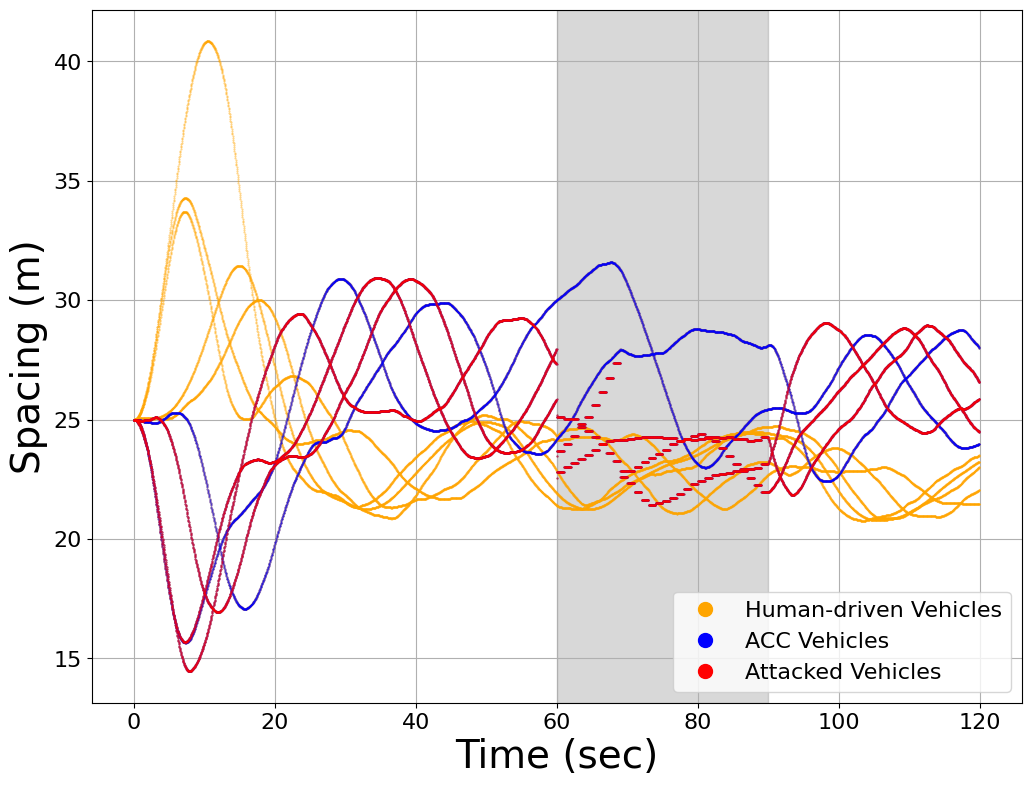}
    \label{ev-spacing-mixed}
}
\hfill
\subfloat[EV speed.]{
    \includegraphics[width=0.45\linewidth]{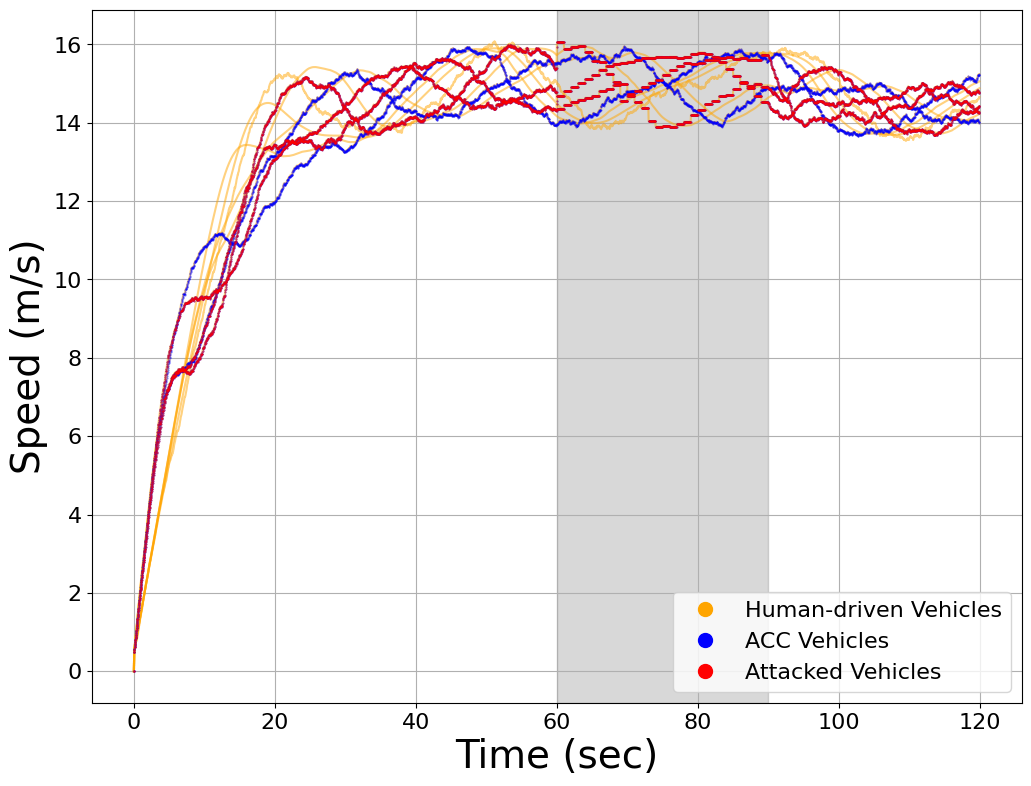}
    \label{ev-speed-mixed}
}

\vspace{0em}

\subfloat[ICE vehicle spacing.]{
    \includegraphics[width=0.45\linewidth]{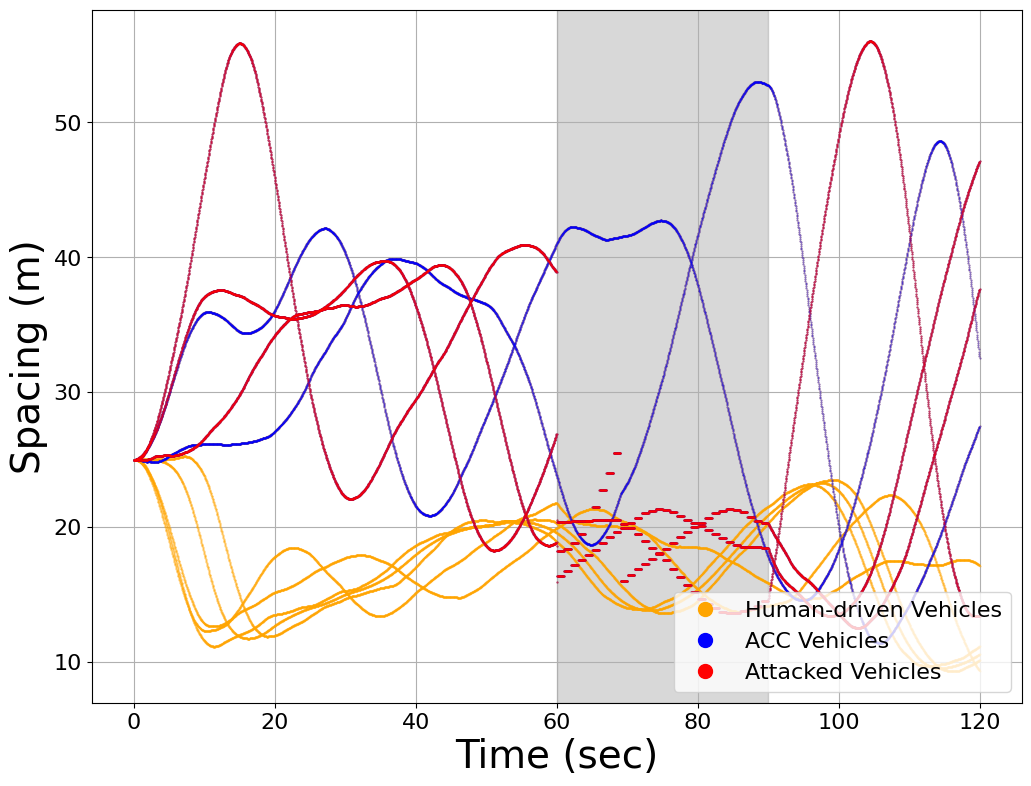}
    \label{ice-spacing-mixed}
}
\hfill
\subfloat[ICE vehicle speed.]{
    \includegraphics[width=0.45\linewidth]{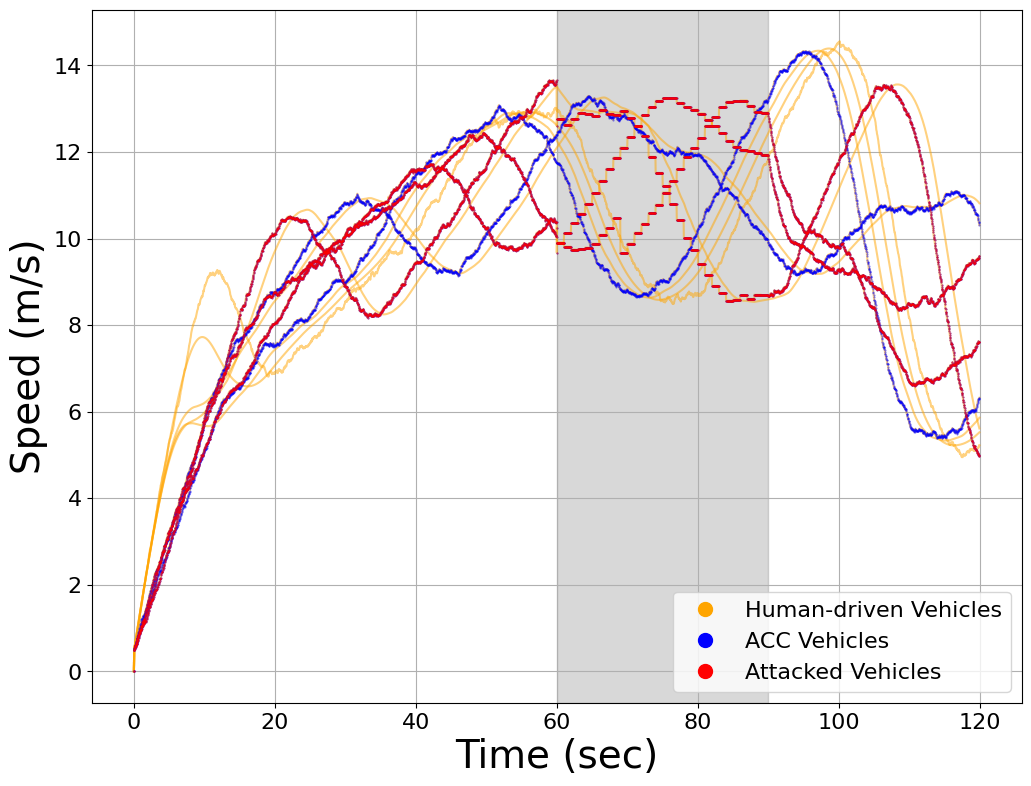}
    \label{ice-speed-mixed}
}

\caption{Comparative analysis of spacing and velocity dynamics between EVs and ICE vehicles under MA in Scenario IV (9-second delay).}
\label{Figuremix attack}
\end{figure}

The comparative analysis in Fig. \ref{Figuremix attack} reveals that EVs maintained stable, high-velocity operation throughout all simulation phases (Fig. \ref{ev-speed-mixed} vs. \ref{ice-speed-mixed}), with minimal spacing variations indicating robust platoon cohesion. In contrast, ICE vehicles exhibited substantial spacing fluctuations (Fig. \ref{ice-spacing-mixed} vs. \ref{ev-spacing-mixed}) and degraded velocity stability, yet remained collision-free. Quantitative analysis in Table \ref{under six cyberattacks} corroborates these observations, with EVs demonstrating superior stability indices with lower $THW$, $\overline{SSD}$, and $\overline{VSD}$ values across all phases. 

\begin{table*}[h]
\caption{\\ Risk-based classification and comparison of proposed cyberattacks on EVs and ICE vehicles based on collision potentials. The symbol ‘---’ indicates the absence of collision events, while the symbol ‘$\times$’ indicates that vehicles consistently experience collisions in all simulation scenarios, thereby precluding post-attack recovery evaluation.}
\label{tab:classification and comparison}
\centering
\resizebox{\textwidth}{!}{
\begin{tabular}{llcccccccccc}
\toprule
\multirow{2}{*}{\textbf{Classification}} & \multirow{2}{*}{\textbf{Attack}} 
& \multicolumn{2}{c}{\textbf{Higher average speed}} 
& \multicolumn{2}{c}{\textbf{Greater velocity stability}} 
& \multicolumn{2}{c}{\textbf{More cohesive platoon}} 
& \multicolumn{2}{c}{\textbf{Shorter time to collision}} 
& \multicolumn{2}{c}{\textbf{Post-attack recovery}} \\
\cmidrule(lr){3-4} \cmidrule(lr){5-6} \cmidrule(lr){7-8} \cmidrule(lr){9-10} \cmidrule(lr){11-12}
& & EV & ICE & EV & ICE & EV & ICE & EV & ICE & EV & ICE \\
\midrule
\multirow{3}{*}{Low-risk attack} 
& PA   & \checkmark &       & \checkmark &       & \checkmark &       & ---         & ---         & \checkmark &        \\
& MA   & \checkmark &       & \checkmark &       & \checkmark &       & ---         & ---         & \checkmark &        \\
& AVA  & \checkmark &       & \checkmark &       & \checkmark &       & ---         & ---         & \checkmark &        \\
\midrule
\multirow{2}{*}{Variable-risk attack} 
& DPDA & \checkmark &       & \checkmark &       & \checkmark &       &             & \checkmark & \checkmark &        \\
& FA   & \checkmark &       & \checkmark &       & \checkmark &       &             & \checkmark & \checkmark &        \\
\midrule
High-risk attack 
& BA   & \checkmark &       & \checkmark &       & \checkmark &       & \checkmark  &            & $\times$   & $\times$ \\
\bottomrule
\end{tabular}
}
\end{table*}

Unexpectedly, MA demonstrated reduced severity compared to its individual components. The DPDA independently precipitated collision events under critical simulation scenarios. However, the combined MA resulted in no collisions across all tested configurations. This counterintuitive outcome persisted even under the most adverse conditions in Scenario IV with adjacent compromised vehicles.

This phenomenon suggests that the interaction between DPDA and PA components may generate compensatory effects that mitigate individual attack impacts. The temporal discretization from DPDA potentially dampens the abrupt behavioral changes induced by PA's falsified information, while the shifted information source from PA may prevent the resonance effects typically associated with DPDA. These findings challenge the prevailing assumption that compound cyberattacks inherently produce additive or synergistic damage, highlighting the complex, nonlinear nature of attack interactions in AV systems.

\subsection{Discussion of Results}\label{sec:summary results}

Through comprehensive modeling and simulation of six distinct cyberattack across varying ACC MPRs and spatial configurations of compromised vehicles, we systematically evaluate their impacts on traffic safety. The results reveal substantial heterogeneity in attack severity, necessitating a risk-based classification framework to guide cybersecurity defense strategies.
Based on collision potential, the proposed cyberattacks are categorized into three risk levels:

\begin{itemize}
    \item \textbf{Low-risk attacks:} PA, MA, and AVA exhibited strong resilience across all simulation scenarios, as evidenced by the absence of collision events. These attacks induced minimal disruption to traffic flow stability and vehicle coordination, thereby maintaining safe operational conditions despite the compromised V2V communication or sensor data integrity.

\item \textbf{Variable-risk attacks:} DPDA and FA exhibit context-dependent severity. Their impact varies significantly based on: (1) the spatial proximity of compromised vehicles, with adjacent configurations demonstrating heightened vulnerability, and (2) attack parameter intensity, such as delay duration for DPDA. Under benign conditions with dispersed attack targets, these attacks maintain stable traffic flow. However, aggressive parameters combined with the configuration of adjacent compromised vehicle can escalate to collision events.

\item \textbf{High-risk attacks:} BA consistently precipitated collision events across all simulation scenarios, regardless of ACC MPRs or vehicle distribution configurations. This attack's mechanism of corrupting spatial perception creates unavoidable failure modes that overwhelm ACC safety margins, warranting immediate mitigation strategies.

\end{itemize}

This risk-based taxonomy is summarized in Table \ref{tab:classification and comparison} which provides a structured framework for prioritizing defense mechanisms, allocating cybersecurity resources, and developing adaptive safety protocols. 
The classification facilitates hierarchical optimization of countermeasures, focusing immediate attention on high-risk threats while implementing graduated responses for addressing variable-risk scenarios. 
Furthermore, this categorization supports regulatory policy development by identifying critical vulnerabilities requiring mandatory protection standards versus optional enhanced security measures.

\section{Conclusions}\label{sec:conclusion}
The proliferation of ACC-equipped AVs, particularly electric platforms, has created an expanded attack surface vulnerable to sophisticated cyberattacks. Malicious actors can exploit these vulnerabilities through communication manipulation, sensor data forgery, and control system compromise, potentially precipitating severe traffic incidents. Despite this growing threat, mathematical frameworks for modeling and analyzing vehicular cyberattacks remain underdeveloped, limiting systematic vulnerability assessment and defense strategy formulation.

This study addresses this critical gap by proposing six novel communication-based cyberattack models targeting ACC-equipped vehicles. 
Through comprehensive simulation across four representative traffic scenarios with varying ACC MPRs and attack configurations, we systematically evaluated the differential impacts on EVs and ICE vehicles. The key findings are summarized as follows.

\textbf{Attack Classification and Risk Assessment:} Based on collision potential, the proposed cyberattacks are classified into three distinct risk categories: (1) low-risk attacks (PA, MA, AVA) that maintain traffic stability and do not induce collisions, (2) high-risk attacks (BA) that consistently result in collision events across all simulation scenarios, and (3) variable-risk attacks (DPDA, FA) whose severity depends on vehicle proximity and attack parameters. In particular, Scenario IV with adjacent attacked vehicles exhibited the highest susceptibility to disruption, whereas Scenario I with isolated attacks showed minimal adverse effects.

\textbf{EV-ICE Performance Differential:} EVs consistently outperformed ICE vehicles across most attack scenarios, maintaining lower $\overline{VSD}$, $\overline{SSD}$, and $THW$ values while achieving higher $V_{avg}$. These metrics collectively indicate that EVs operate with tighter platoon formations, more stable velocities, and superior efficiency. However, this aggressive driving profile paradoxically increases EV vulnerability to BA, where corrupted spatial perception transforms their operational advantages into safety liabilities—EVs experienced collisions 25.8 seconds earlier than ICE vehicles under BA conditions.

\textbf{Post-Attack Recovery Dynamics:} The post-attack phase revealed fundamental disparities in system resilience of two vehicle types. EVs demonstrated rapid stabilization with consistently lower $\overline{VSD}$ and $\overline{SSD}$ values, indicating superior disturbance rejection and recovery capabilities. In contrast, ICE vehicles exhibited prolonged instability with elevated fluctuation metrics. This performance gap stems from EVs' inherent advantages: instantaneous torque response, advanced control algorithms, and integrated sensor-actuator architectures that enable rapid adaptation to dynamic driving conditions.

\textbf{Implications for Cybersecurity:} The context-dependent nature of cyberattack impacts necessitates adaptive defense strategies. While EVs' superior dynamics provide resilience against most attacks, their vulnerability to perception-based attacks requires specialized countermeasures. Notably, the counterintuitive finding that MA yields less severe outcomes than its individual components suggests complex, nonlinear attack interactions that warrant further investigation.

Future research should extend this framework in several directions: (1) developing real-time detection algorithms for the proposed attack signatures, (2) designing attack-specific mitigation strategies that account for powertrain-dependent vulnerabilities, (3) investigating the scalability and propagation dynamics of cyberattacks in high-density traffic scenarios, and (4) exploring the potential for cascading failures in mixed-fleet environments. Additionally, the integration of machine learning techniques for anomaly detection and the development of resilient ACC algorithms that maintain safety under compromised conditions represent critical areas for advancing AV cybersecurity.

\newpage
\bibliographystyle{IEEEtran}
\bibliography{refs}

\end{document}